\documentclass{vldb}

\usepackage{times}
\usepackage{graphicx}
\usepackage{amsmath}
\usepackage{balance}
\usepackage{url}
\usepackage{comment}
\usepackage{enumerate}
\usepackage{lipsum}
\usepackage{enumitem}
\usepackage[colorinlistoftodos, textwidth=4cm, shadow, backgroundcolor=yellow]{todonotes}
\usepackage{subfigure}
\usepackage{chngpage}
\usepackage{calc}
\usepackage{pbox}
\usepackage{array}

\newcommand{\compactlist}{
  \begin{list}{$\bullet$}{
    \setlength{\itemsep}{0pt}
    \setlength{\parsep}{3pt}
    \setlength{\topsep}{3pt}
    \setlength{\partopsep}{0pt}
    \setlength{\leftmargin}{1.5em}
    \setlength{\labelwidth}{1em}
    \setlength{\labelsep}{0.5em}
  }
}
\newcommand{\compactend}{\end{list}}

\usepackage{paralist}

\begin{document}

\title{S-Store: Streaming Meets Transaction Processing}

\numberofauthors{1}
\author{
  \alignauthor
    John Meehan$^{1}$, Nesime Tatbul$^{2,3}$, Stan Zdonik$^{1}$, Cansu Aslantas$^{1}$,  \\
    Ugur Cetintemel$^{1}$,Jiang Du$^{4}$,Tim Kraska$^{1}$,Samuel Madden$^{3}$,David Maier$^{5}$,\\
    Andrew Pavlo$^{6}$,Michael Stonebraker$^{3}$,Kristin Tufte$^{5}$,Hao Wang$^{3}$ \\
    ~ \\
  \affaddr{
    $^{1}$Brown University \hbox{~~~~}
    $^{2}$Intel Labs \hbox{~~~~}
    $^{3}$MIT \hbox{~~~~}
    $^{4}$UToronto \hbox{~~~~}
    $^{5}$Portland State University \hbox{~~~~}
    $^{6}$CMU
  }
}

\maketitle

\begin{abstract}

Stream processing addresses the needs of real-time applications. Transaction processing addresses the coordination and safety of short atomic computations. Heretofore, these two modes of operation existed in separate, stove-piped systems. In this work, we attempt to fuse the two computational paradigms in a single system called S-Store. In this way, S-Store can simultaneously accommodate OLTP and streaming applications. We present a simple transaction model for streams that integrates seamlessly with a traditional OLTP system. We chose to build S-Store as an extension of H-Store, an open-source, in-memory, distributed OLTP database system. By implementing S-Store in this way, we can make use of the transaction processing facilities that H-Store already supports, and we can concentrate on the additional implementation features that are needed to support streaming. Similar implementations could be done using other main-memory OLTP platforms. We show that we can actually achieve higher throughput for streaming workloads in S-Store than an equivalent deployment in H-Store alone. We also show how this can be achieved within H-Store with the addition of a modest amount of new functionality. Furthermore, we compare S-Store to two state-of-the-art streaming systems, Spark Streaming and Storm, and show how S-Store matches and sometimes exceeds their performance while providing stronger transactional guarantees.

\end{abstract}

\section{Introduction} \label{sec:intro}

A decade ago, the database research community focused attention on stream data processing systems. These systems \cite{stream, telegraphcq}, including our own system, Aurora/Borealis \cite{aurora, borealis}, were largely concerned with executing SQL-like operators on an unbounded and continuous stream of input data. The  main optimization goal of these systems was reducing the latency of results, since they mainly addressed what might be called monitoring applications \cite{lerner2003virtues, johnson2008query}. To achieve this, they were typically run in main memory, thereby avoiding the latency-killer of the disk.

While essentially all of the monitoring applications that we encountered had a need for archival storage, the system-level support for this was limited and ad hoc. That is, the systems were largely not designed with storage in mind; it was tacked on after the fact. Thus, there was no support for things like ACID transactions, leaving applications open to potential inconsistencies with weak guarantees for isolation and recovery semantics.

These first-generation streaming systems could be viewed as real-time analytics systems. After all, the input was made up of an infinite stream of new tuples. The notion of some of these tuples representing updates of previously viewed tuples was not made explicit in the model. This is fine if time is the key. In this case, if each tuple is given a unique timestamp, the update pattern is append-only. However, there are cases when the identifying attribute is something else. Consider a stock ticker application in which stock symbol is the key.  Here a new tuple for, say, IBM is really an update to the previously reported price. Traders want to see the current stock book as a consistent view of the 6000 stocks on the NYSE, with all prices reported in a consistent way.

We are beginning to see the rise of a second generation of streaming systems \cite{storm, zaharia13}. These systems do not enforce a relational view on their users. Instead, they allow users to create their own operators that are invoked and managed by a common infrastructure. Note that it is reasonable to have libraries of common operators, including relational operators, that manipulate tables. The infrastructure enforces some model of failure semantics (e.g., at-least-once or exactly-once processing), but still ignores needs of proper isolation and consistent storage in the context of updates.

Meanwhile, the advent of inexpensive, high-density RAM has led to a new-generation of distributed on-line transaction processing (OLTP) systems that store their data in main memory, thereby enabling very high throughput with ACID guarantees (e.g., \cite{voltdb, hekaton, hstore}). However, these systems lack the notion of stream-based processing (e.g., unbounded data, push-based data arrival, ordered processing, windowing).

Many applications need aspects of both streaming and transaction processing. In this paper, we propose to combine these two computational paradigms under a single system called S-Store. Applications that benefit from this kind of a hybrid system include those that use the streaming facilities to record persistent state or views in shared tables, and at the same time use the OLTP facilities to display a representation or summary of this state \cite{sstore-demo}. Obviously, one would want the displayed representation to show a consistent state of the world. Examples here include dashboards with multiple summary displays or leaderboards that show aggregate statistics derived from the shared persistent state. We describe a detailed example below.

\subsection{A Motivating Example} \label{sec:example}

One application domain that benefits greatly from a hybrid system such as S-Store is leaderboard maintenance. For example, consider an American Idol-like TV show, in which a number of candidates are presented to viewers for them to vote for the candidate that they find the most appealing. Each viewer may cast a single vote via text message. Suppose that a preferential voting scheme is used, where the candidate with the fewest number of votes will be removed from the running every 1000 votes, as it has become clear that s/he is the least popular. When this candidate is removed, votes submitted for him or her will be deleted, effectively returning the votes to the people who cast them. Those votes may then be re-submitted for any of the remaining candidates. This continues until a single winner is declared. During the course of the voting, each incoming vote needs to be validated and recorded. Furthermore, several leaderboards are maintained: one representing the top-3 candidates, another for the bottom-3 candidates, and a third one for the top-3 trending candidates of the last 100 votes. With each incoming vote, these leaderboards are updated with new statistics regarding the number of votes each candidate has received.

\begin{figure}[t]
\centering \includegraphics[width=\columnwidth]{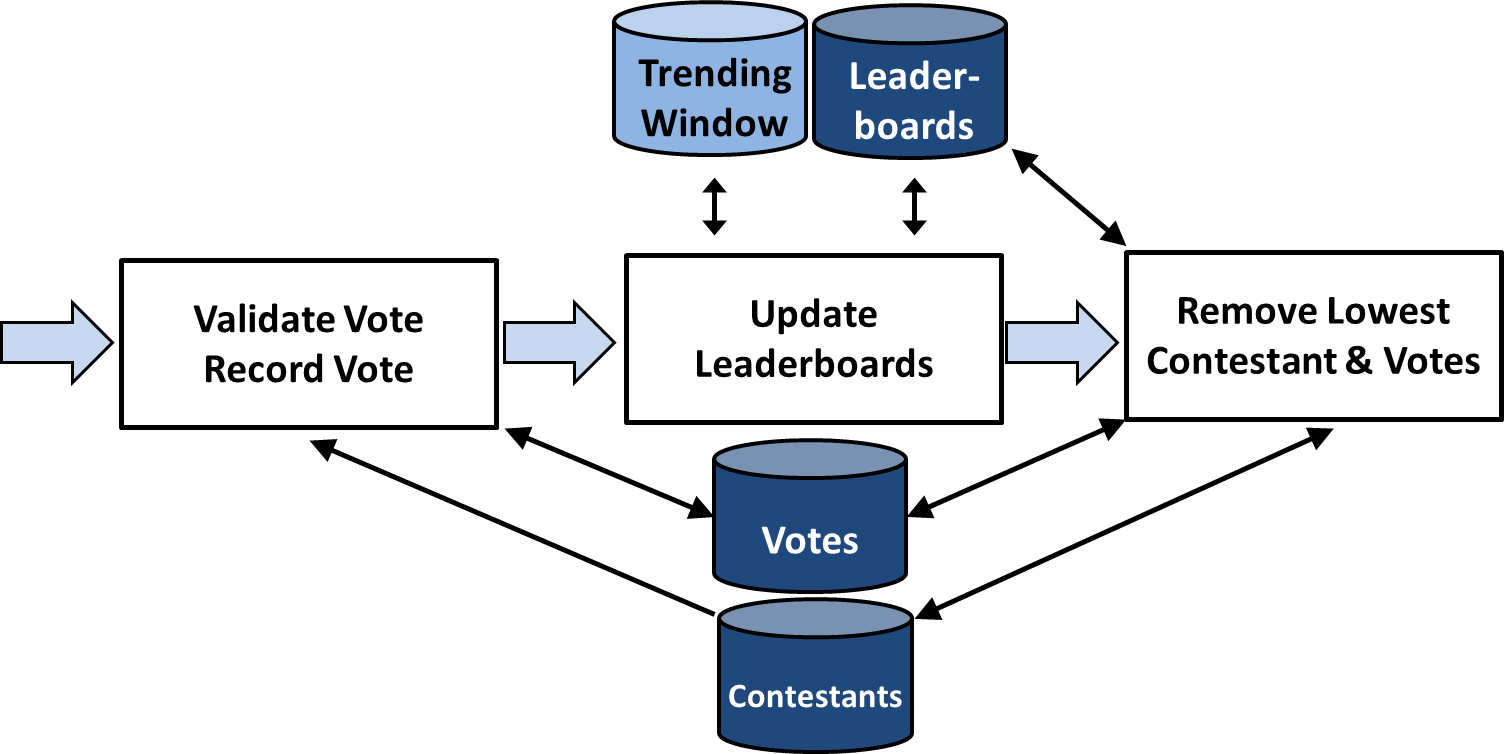}
\caption{Leaderboard Maintenance Workflow}
\vspace{-0.2in}
\label{f:LB-workflow}
\end{figure}

One can model this application with a workflow that consists of three logical transactional steps, as shown in Figure \ref{f:LB-workflow}: (i) one to validate new votes (i.e., check that the corresponding contestant exists, check that the corresponding viewer has not voted before, etc.) and record them; (ii) another to maintain the leaderboards; and (iii) a third to remove the lowest contestants and votes every 1000 votes, also updating the relevant leaderboards as necessary.

This workflow requires both stream-based processing and transactional state management. Votes continuously arrive as a stream, and the Votes and Contestants tables keep a record of all previously received valid votes together with the viewers who submitted them. Each vote's validation requires a read of the Votes and Contestants tables, followed by a write to the Votes table to record the newly validated vote. The leaderboard maintenance transaction maintains a sliding window over the most recent 100 valid votes in order to aggregate them into a count for trend computation. Additionally, it must maintain a total vote count for each running contestant in order to identify the top and bottom performers. Finally, the last transaction will trigger every 1000 incoming votes to update the Votes, Contestants, and Leaderboard tables for the removed candidate. State is shared among the three transactions, and all state accessed during this workflow (i.e., Votes, Contestants, Leaderboards, Trending Window) must be kept consistent.   In fact, the three transactions in this workflow must always run serially for each incoming vote in order to ensure correct behavior. Otherwise, votes might be validated/invalidated incorrectly and the wrong contestants might be removed, eventually leading to the wrong winner. Last but not least, all state must be maintained persistently and recovered correctly (i.e., in an ACID manner) in the event of a failure.

\subsection{Contributions and Outline} \label{sec:outline}

This paper introduces the design and implementation of S-Store, a single system for processing streams and transactions with well-defined correctness guarantees. Our approach to building such a system is to start with a fully transactional OLTP main-memory database system and to integrate additional streaming functionality on top of it. By doing so, we are able to leverage infrastructure that already addresses many of the implementation complexities of transaction processing. Also, this choice is very natural, since streaming systems largely run in main memory to achieve low latency. More specifically, this work makes the following key contributions:

\noindent
{\bf Model.}
We define a novel, general-purpose computational model that allows us to seamlessly mix {\em streaming transactions} with ordinary OLTP transactions. Stream processing adds additional semantics to an OLTP engine's operating model. In particular, stream processing introduces the notion of {\em order} to the transaction mix. That is, it is possible to say that one transaction must precede another, something that is missing from the non-deterministic semantics of a standard transaction model. Further, since streams are unbounded and arrive on a continuous basis, there is a need to add the necessary primitives for bounding computation on streams such as batch-based processing \cite{stream, streamsql} and windowing \cite{golab-sigrec03, secret}. Finally, streaming transactions support a push-based processing model, whereas OLTP transactions access state in a pull-based manner. Our hybrid model provides a uniform way to represent and access state for all transactions.

\noindent
{\bf Architecture and Implementation.}
We show how such a hybrid computational model can be cleanly and efficiently implemented on top of a state-of-the-art main-memory OLTP engine. While we have designed and implemented S-Store on top of H-Store \cite{hstore}, our architectural extensions are general enough to be applied to any main-memory OLTP engine. These include: (i) streams and windows represented as time-varying state, (ii) triggers to enable push-based processing over streams and windows, (iii) a streaming scheduler that ensures correct transaction ordering, and (iv) a variant on H-Store's recovery scheme \cite{malviya14} that is more appropriate for stream processing. Please note that the discussion in this paper is confined to the single-node case; multi-node S-Store is the topic for a follow-on paper.

\noindent
{\bf Performance.}
We provide a detailed study of S-Store's performance characteristics, including the surprising observation that S-Store can improve transaction throughput over H-Store \cite{hstore}, as well as outperforming Spark Streaming \cite{zaharia13} and Storm \cite{storm}, on a number of benchmarks including the leaderboard example shown in Figure \ref{f:LB-workflow}.

The rest of this paper is organized as follows: We first describe our computational model for transactional stream processing in Section \ref{sec:model}. Section \ref{sec:implementation} presents the design and implementation of the S-Store system, which realizes this model on top of the H-Store main-memory OLTP system \cite{hstore}. In Section \ref{sec:experiments}, we present an experimental evaluation of S-Store in comparison to H-Store as well as to two state-of-the-art stream processing systems - Spark Streaming \cite{zaharia13} and Storm \cite{storm}. We discuss related work in Section \ref{sec:related}, and finally conclude the paper with a summary and a discussion of future research directions in Section \ref{sec:summary}.

\section{The Computational Model} \label{sec:model}

In this section, we describe our computational model for transactional stream processing. This model inherits traditional, well-accepted models of OLTP and stream processing, and fuses them within one coherent model which will then allow us to support hybrid workloads (independent OLTP transactions and streaming transactions) with well-defined transactional guarantees.

We now describe our overall abstract models of OLTP and streaming, and how we blended them together. Our goal is to design S-Store in such a way that ordinary OLTP transactions can operate concurrently with streaming transactions. We assume that both kinds of transactions can share state and at the same time produce correct results. We assume three different kinds of state: (i) public shared tables, (ii) windows, and (iii) streams. Furthermore, we make a distinction between {\em OLTP transactions} (only access (i)) and {\em streaming transactions} (access (iii) + optionally (ii) or (i)).

For OLTP transactions, we simply adopt the traditional ACID model that has been well-described in the literature \cite{weikum-book}. A {\em database} consists of unordered, bounded collections (i.e., sets) of tuples. A {\em transaction} represents a finite unit of work (i.e., a finite sequence of read and write operations) performed over a given database. Each transaction has a definition and possibly many executions (i.e., instances). We assume that all transactions are predefined as {\em stored procedures} with input parameters, each of which can be instantiated with specific parameter values through explicit client invocations (``pull"). In order to maintain integrity of the database in the face of concurrent transaction executions and failures, each transaction is executed with {\em ACID guarantees}.

Streaming transactions are also predefined as parametric stored procedures, but they are instantiated rather differently, i.e. as new input data becomes available (``push"). We describe streaming transactions in more detail next.

\subsection{Streaming Transactions and Workflows} \label{sec:streaming}

Our stream data model is very similar to many of the stream processing systems of a decade ago \cite{aurora, stream, telegraphcq}. A {\em stream} is an ordered, unbounded collection of tuples. Tuples have a timestamp \cite{stream} or batch-id \cite{streamsql, secret} that specifies simultaneity. Tuples with the same timestamp or batch-id logically occur as a group at the same time and, thus, should be processed as a unit.

A {\em window} is a finite, contiguous subsequence of a stream. Windows can be defined in many different ways \cite{golab-sigrec03, secret}, but for the purposes of this work, we will restrict our focus to the most common type - sliding windows. A {\em sliding window} is a window which has a fixed size and a fixed slide, where the slide specifies the distance between two consecutive windows and can be less than or equal to size (called a {\em tumbling window} if the latter). A sliding window is said to be {\em time-based} if its size and slide are defined in terms of tuple timestamps, and {\em tuple-based} if its size and slide are defined in terms of the number of tuples.

Stream processing systems commonly define computations over streams as workflows. Early streaming systems focused on relational-style operators as computations (e.g., Filter, Join), whereas current systems support more general user-defined computations \cite{storm, zaharia13}. Following this trend and consistent with our OLTP model, we assume that computations over streams are expressed as workflows of user-defined stored procedures. More formally, a {\em workflow} is a directed acyclic graph (DAG), in which nodes represent streaming transactions and edges represent an execution ordering. If there is an edge between node $N_i$ and node $N_j$, there is also a stream that is output for $N_i$ and input for $N_j$. We say $N_i$ precedes $N_j$ and is denoted as $N_i \prec N_j$.

Given the unbounded nature of streams, we restrict streaming transaction instances to operate over non-overlapping ``atomic batch\-es" of stream tuples. An {\em atomic batch} is a finite, contiguous subsequence of a stream that must be processed atomically  (e.g., tuples with the same timestamp, from a common time interval, that belong to the same sales order, etc.).

Contiguous atomic batches of tuples arrive on a stream at the input to a workflow from push-based data sources. We adopt the data-driven execution model of streams, where arrival of a new atomic batch causes a new invocation for all the streaming transactions that are defined over the corresponding stream. This is in contrast to the on-demand execution model of OLTP transactions which are invoked by explicit requests from clients.

\begin{figure}[t]
\centering
\includegraphics[width=0.8\columnwidth]{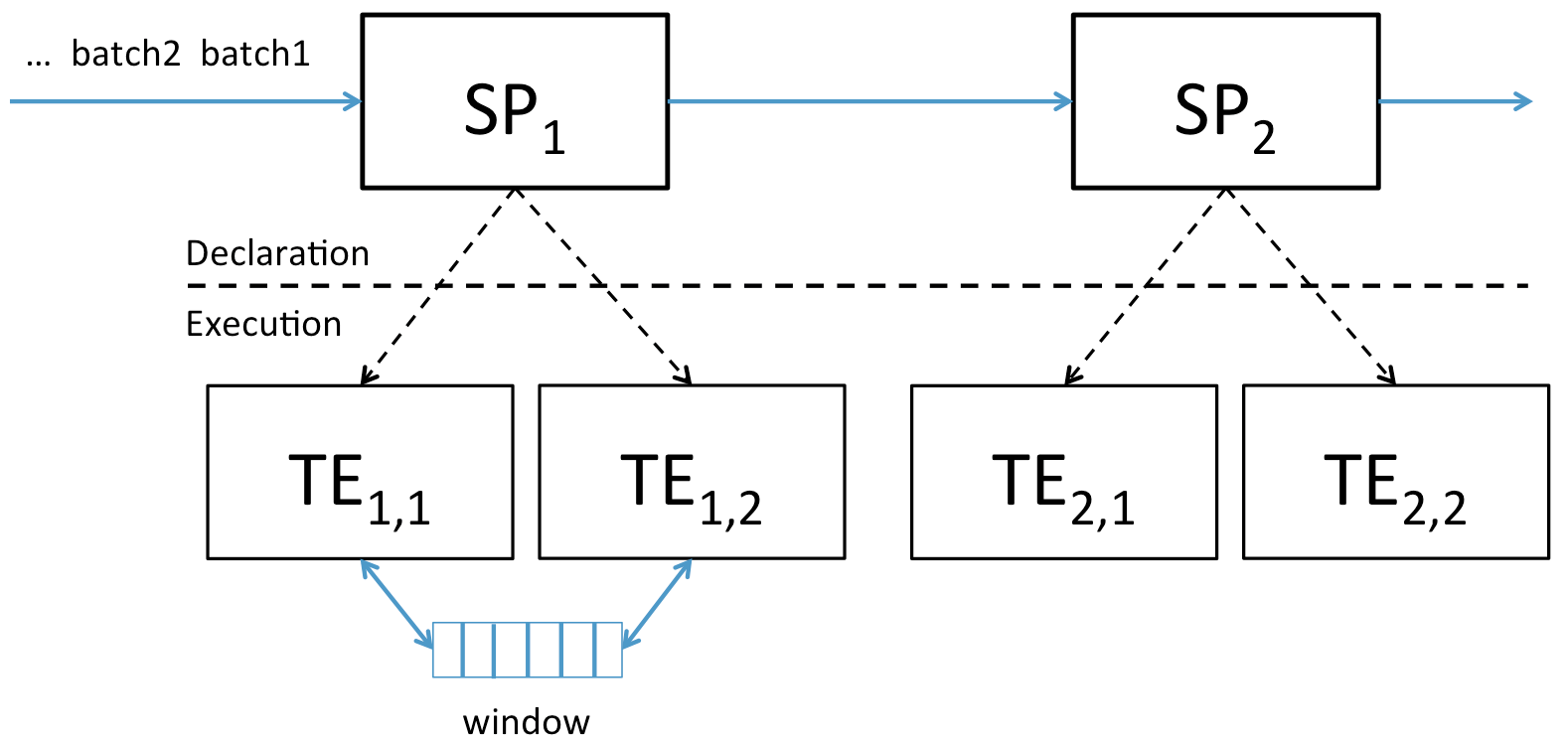}
\caption{Transaction Executions in a Workflow}
\vspace{-0.2in}
\label{f:workflow}
\end{figure}

Streaming transactions are defined as stored procedures once, but they are executed many times, once for each input batch. We refer to each such execution of a transaction definition as a {\em transaction execution} (TE). A TE essentially corresponds to an atomic batch and its subsequent processing by a stored procedure. For example, in Figure \ref{f:workflow}, a workflow with two stored procedures are defined, but each of those are executed twice for two contiguous atomic batches, yielding a total of four transaction executions.

Given a workflow, it is also useful to distinguish between {\em border transactions} (those that ingest streams from the outside, e.g., $SP_1$ in Figure \ref{f:workflow}) and {\em interior stored procedures} (others, e.g., $SP_2$ in Figure \ref{f:workflow}). Border stored procedures are instantiated by each newly arriving atomic batch, and each such execution may produce a group of output stream tuples labeled with the same batch-id as the input that produced them. These output tuples become the atomic batch for the immediately downstream interior stored procedures, and so on.


In the rest of this paper, we use the terms ``transaction'' and ``stored procedure'' interchangeably to refer to the definition of a transaction, whereas ``transaction execution'' to refer to a specific instance of that definition.

\subsection{Correct Execution} \label{sec:isolation}

A standard OLTP transaction mechanism guarantees the isolation of one transaction's operations from other transactions'. When a transaction $T$ commits successfully, all of $T$'s writes are installed and made public. During the execution of $T$, all of $T$'s writes remain private.

S-Store adopts such standard transaction semantics as a basic building block for its streaming transactions; however, the ordering of stored procedures in the workflow definition as well as the inherent order in streaming data puts additional constraints on allowable transaction execution orders. As an example, consider the workflow shown in Figure \ref{f:workflow}, where $SP_1$ precedes $SP_2$. We use $TE_{i,j}$ to indicate the $j$th transaction execution of stored procedure $SP_i$. The figure illustrates four transaction executions that correspond to the processing of two atomic input batches (batch1 and batch2) through the complete workflow. These TE's can be ordered in one of two possible ways:
$[TE_{1,1}, TE_{2,1}, TE_{1,2}, TE_{2,2}]$ or $[TE_{1,1}, TE_{1,2}, TE_{2,1}, TE_{2,2}]$. Any other orderings would not lead to a correct execution. This is in contrast to most OLTP transaction processors which would accept any serializable schedule (e.g., one that is equivalent to any of the 4! possible serial execution schedules if these were 4 independent OLTP transactions).

Note that we make no ACID claims for the workflow as a whole. The result of running a workflow is to create an ordered execution of ACID transactions.

Furthermore, in streaming applications, the state of a window must be shared differently than other stored state. To understand this, consider again the simple workflow shown in Figure \ref{f:workflow}. Let us assume for simplicity that the transaction input batch size for $SP_1$ is one tuple. Further, suppose that $SP_1$ constructs a window of size 2 that slides by 1 tuple, i.e., two consecutive windows in $SP_1$ overlap by 1 tuple. This means that window state will carry over from $TE_{1,1}$ to $TE_{1,2}$. For correct behavior, this window state must not be publicly shared with other TE's. That is, the state of a window can be shared among transaction executions of a given stored procedure, but should not be made public beyond that. Returning to Figure \ref{f:workflow}, when $TE_{1,1}$ commits, the window in $TE_{1,1}$ will slide by one and will then be available to $TE_{1,2}$, but not to $TE_{2,1}$. This approach to window visibility is necessary because of the nature of streaming applications. In some sense, it is this way of sharing window state that is the basis for continuous operation. Windows evolve and, in some sense, ``belong'' to a particular stored procedure. Thus, a window's visibility should be restricted to the transaction executions of its owning stored procedure.

We will now describe what constitutes a {\em correct execution} for a workflow of streaming transactions more formally. Consider a workflow $W$ of $n$ streaming transactions $T_i$, $1 \leq i \leq n$. $W$ is a directed acyclic graph $G=(V,E)$, where $V=\{T_1, \ldots, T_n\}$ and $E \subseteq V \times V$, where $(T_i, T_j) \in E$ means that $T_i$ must precede $T_j$ (denoted as $T_i \prec T_j$). Being a DAG, $G$ has at least one topological ordering. A topological ordering of $G$ is an ordering of its nodes $T_i \in V$ such that for every edge $(T_i, T_j) \in E$ we have $i < j$. Each topological ordering of $G$ is essentially some permutation of $V$.

Without loss of generality:
(i) Let us focus on one specific topological ordering of $G$ and call it $O$;
(ii) For ease of notation, let us simply assume that $O$ corresponds to the identity permutation such that it represents: $T_1 \prec T_2 \prec .. \prec T_n$.

$T_i$ represents a transaction definition $T_i(s_i, w_i, p_i)$, where
$s_i$ denotes all private stream inputs of $T_i$ (at least one),
$w_i$ denotes all private window inputs of $T_i$ (optional),
$p_i$ denotes all public table partition inputs of $T_i$ (optional).
Similarly, $T_i^j$ represents the $j^{\text{th}}$ transaction execution of $T_i$ as $T_i^j(s_i.b_j, w_i, p_i)$, where $b_j$ denotes the $j^{\text{th}}$ atomic batches of all streams in $s_i$.

A workflow $W$ is executed in rounds of atomic batches $1 \leq r < \infty$, such that for any round $r$, atomic batch $r$ from all streaming inputs into $W$ generates a sequence of TE's $T_i^r(s_i.b_r, w_i, p_i)$ for each $T_i$.
Note that this execution generates an {\em unbounded schedule}. However, as of a specific round $r=R$, we generate a {\em bounded schedule} that consists of all $R*n$ TE's: $1 \leq r \leq R, 1 \leq i \leq n, T_i^r(s_i.b_r, w_i, p_i)$.

In the traditional ACID model of databases, any permutation of these $R*n$ TE's would be considered to be a valid/correct, serial schedule. In our model, we additionally have:

\begin{compactenum}

\item {\em Workflow order constraint:}
Consider the topological ordering $O$ of $W$ as we defined above. Then for any given execution round $r$, it must hold that:

$T_1^r(s_1.b_r, w_1, p_1) \prec \ldots \prec T_n^r(s_n.b_r, w_n, p_n)$ 

\item {\em Stream order constraint:}
For any given transaction $T_i$, as of any round $r=R$, the following must hold:

$T_i^1(s_i.b_1, w_i, p_i) \prec \ldots \prec T_i^R(s_i.b_R, w_i, p_i)$

\end{compactenum}

Any bounded schedule of $W$ that meets the above two ordering constraints is a {\em correct schedule}.

If $G$ has multiple topological orderings, then the workflow order constaint must be relaxed to accept any of those orderings for any given execution round of $W$.

\subsection{Hybrid Transaction Schedules} \label{sec:hybrid}

S-Store's computational model allows OLTP and streaming transactions to co-exist as part of a common transaction execution schedule. This is particularly interesting if those transactions access public shared tables.

Given previous section's formal description of a correct schedule for a workflow $W$ that consists of streaming transactions, any OLTP transaction execution $T_i^j(p_i)$ (defined on one or more public table partitions $p_i$) is allowed to interleave {\em anywhere} in such a schedule. The resulting schedule would still be correct.

\begin{figure}[t]
\centering
\includegraphics[width=0.4\columnwidth]{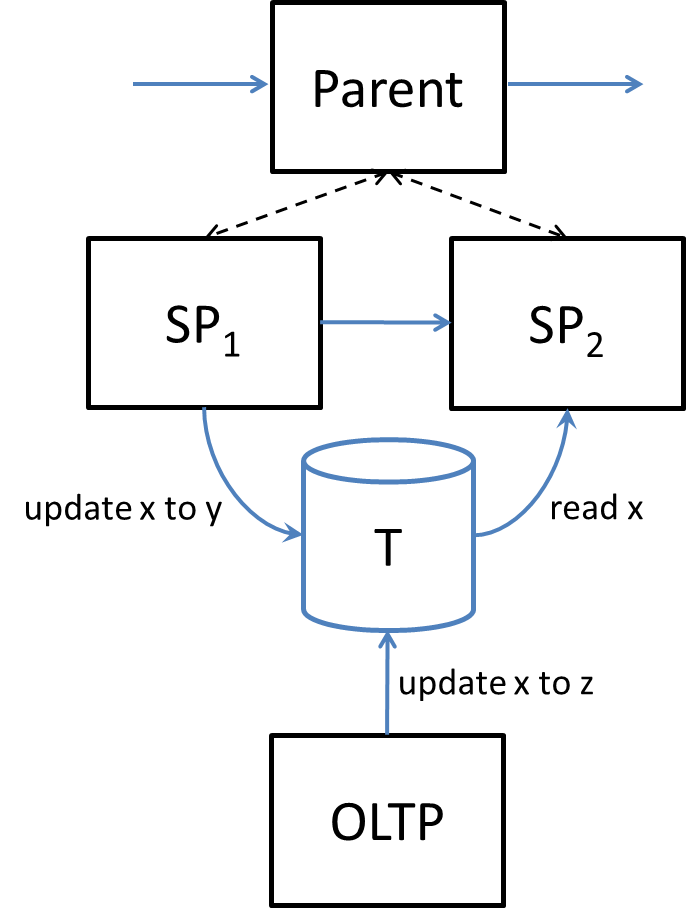}
\caption{Nested Transaction Example}
\vspace{-0.2in}
\label{f:OLTP}
\end{figure}

We have also extended our transaction model to include {\em nested transactions}. Fundamentally, this allows the application programmer to build higher-level transactions out of smaller ones, giving him/her the ability to create coarser isolation units among stored procedures, as illustrated in Figure \ref{f:OLTP}. In this example, two streaming transactions, $SP_1$ and $SP_2$, in a workflow access a shared table $T$. $SP_1$ writes to the table and $SP_2$ reads from it. If another OLTP transaction also writes to $T$ in a way to interleave between $SP_1$ and $SP_2$, then $SP_2$ may get unexpected results. Creating a nested transaction with $SP_1$ and $SP_2$ as its children will isolate the behavior of $SP_1$ and $SP_2$ as a group from other transactions (i.e., other OLTP or streaming). Note that nested transactions can also be useful for isolating multiple instances of a given streaming workflow from one another. For example, defining our leaderboard maintenance workflow described in Section \ref{sec:example} as a nested transaction would facilitate enforcing the correct execution ordering.

More generally, an S-Store nested transaction consists of two or more stored procedures with a partial order defined among them \cite{db-book}. The stored procedures within a nested transaction must execute in a way that is consistent with that partial order.
A nested transaction will commit, if and only if all of its stored procedures commit. If one or more stored procedures abort, the whole nested transaction will abort.

Nested transactions fit into our formal model of streaming transactions in a rather straight-forward way. More specifically, any streaming transaction $T_i$ in workflow $W$ can be defined as a nested transaction that consists of children $T_{i1}, \ldots, T_{im}$. In this case, $T_{i1}, \ldots, T_{im}$ must obey the partial order defined for $T_i$ for every execution round $r$, $1 \leq r < \infty$. This means that no other streaming or OLTP transaction instance will be allowed to interleave with $T_{i1}, \ldots, T_{im}$ for any given execution round.

\subsection{Fault Tolerance} \label{sec:fault}

In the face of failure, S-Store must recover its state such that any committed transactions remain stable and at the same time, any uncommitted transactions are not allowed to have any effect on the publicly shared tables. For a streaming workflow, if $TE_{i,j}$ precedes $TE_{m,n}$, then $TE_{i,j}$ must commit before $TE_{m,n}$. Also, if $TE_{i,j}$ precedes $TE_{i,j+1}$ and they share a window $w$, $w$ must be restored to the state that it had before $TE_{i,j+1}$ began executing. A TE that had started, but had not committed should be undone and it should be restarted with the proper stream input from its predecessor.

S-Store strives to produce {\em exactly-once} executions which is to say that the output will not contain any duplicates, but will produce all output tuples that would have been produced if there were no failures (no lost results). However, if the workflow definition allows multiple transaction execution orderings or if the transactions within a workflow contain any non-deterministic operations, we provide an additional recovery option that we call {\em weak recovery}. Weak recovery will produce a correct result in the sense that it will produce results that could have been produced before the failure, but not necessarily the one that was in fact being produced.

We also are willing to accept a mechanism that is slow on recovery if it is fast during normal operation. This is based on the observation that we expect failures to be rare. In the next section, we will explain the recovery mechanisms that S-Store uses to implement this fault tolerance model.

\section{Architecture \& Implementation} \label{sec:implementation}

We chose to build S-Store on top of the H-Store OLTP system \cite{hstore}. This allows us to inherit H-Store's support for high-throughput transaction processing, thereby eliminating the need to replicate this complex functionality. We also get important associated functionality like indexing, main-memory operation, and support for user-defined transactions that will be important for streaming OLTP applications.

In this section, we describe the H-Store architecture and the chan\-ges that were required to incorporate S-Store's hybrid computational model described in the previous section. An important design goal was to make use of H-Store's transaction model as much as possible, making as few changes and additions to the code-base as possible. Nevertheless, we believe that the architectural features that we have added to H-Store are conceptually applicable to any main-memory OLTP system in general.

\subsection{H-Store Overview} \label{sec:hstore}

H-Store is an open-source, main-memory OLTP engine that was developed at Brown and MIT \cite{hstore}, and formed the basis for the design of the VoltDB NewSQL database system \cite{voltdb}.

All transactions in H-Store must be predefined as stored procedures with input parameters.
The stored procedure code is a mixture of SQL and Java.
Transaction executions are instantiated by binding input parameters of a stored procedure to real values and running it. In general, a given stored procedure will, over time,  generate many transaction executions. Transaction executions are submitted to H-Store, and the H-Store scheduler executes them in whatever order it deems best in order to provide ACID guarantees.

H-Store follows a typical distributed architecture in which there is a layer (called {\em the partion engine (PE)}) that is responsible for managing transaction distribution, scheduling, coordination, and recovery. The PE itself is built on top of another layer (called {\em the execution engine (EE)}) that is responsible for the local execution of SQL queries.
A client program connects to the PE via a stored procedure execution request. If the stored procedure requires SQL processing, then the EE is invoked with these sub-requests.

An H-Store database is partitioned across multiple sites \cite{pavlo12}. A transaction is executed on the sites that hold the data that it needs. If the data is partitioned carefully, most transactions will only need data from a single site. Single-sited transactions are run serially on that site, thereby eliminating the need for fine-grained locks and latches.

H-Store provides recovery through a checkpointing and command-logging mechanism \cite{malviya14}. Periodically, the system creates a persistent snapshot or checkpoint of the current committed state of the database.
Furthermore,
every time H-Store commits a transaction, it writes a command-log record containing the name of the corresponding stored procedure along with its input parameters. This command-log record must be made persistent before its transaction can commit. In order to minimize interactions with the slow persistent store, H-Store offers a group-commit mechanism.


On recovery, the system's state is restored to the latest snapshot, and the command-log is replayed. That is, each command-log record allows the system to re-execute its associated stored procedure with the same arguments that it did before the failure. Note that an undo-log is unnecessary, since the previous checkpoint will not contain uncommitted changes.


\subsection{S-Store Extensions} \label{sec:arch}

\begin{figure*}[t]
\centering
\includegraphics[width=0.7\textwidth]{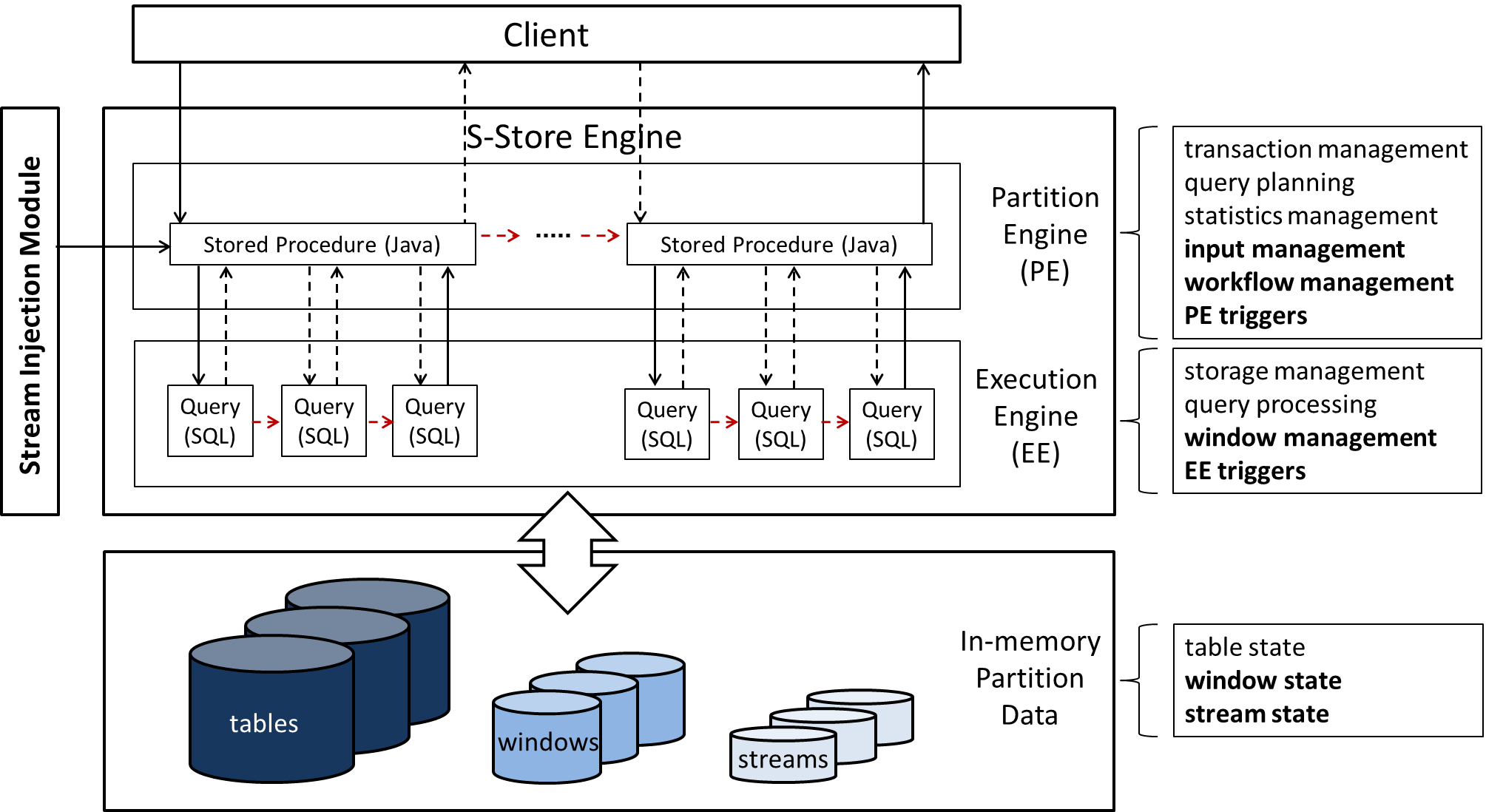}
\caption{S-Store Architecture}
\vspace{-0.2in}
\label{f:architecture}
\end{figure*}

The high-level architecture of S-Store, directly adopted from H-Store, is shown in Figure \ref{f:architecture} and consists of four main components:

\begin{compactitem}

\item a \textbf{client} process that invokes stored procedures with their correct input parameters. This is essentially where an S-Store application runs.

\item a \textbf{stream injection module} that feeds streaming inputs into the S-Store engine. This module is responsible for managing all issues related to input handling such as preparing the atomic batches.

\item a \textbf{partition engine (PE)} that is responsible for breaking the client request into pieces that can be executed at the appropriate partition(s). There is one S-Store partition per core. The available DRAM for a node is divided equally among the partitions, and each stores a horizontal slice of the database.

\item an \textbf{execution engine (EE)} that evaluates SQL expressions. This is called by the partition engine whenever a stored procedure contains SQL.

\end{compactitem}

S-Store makes a number of extensions to H-Store to enable stream processing in the engine (shown in boldface in Figure \ref{f:architecture}). These include: (i) management of inputs from streaming clients and complex workflows of stored procedures at the PE layer, (ii) management of stream- and window-based queries at the EE layer, (iii) management of in-memory stream and window state.
Furthermore, S-Store makes critical extensions to transaction scheduling and recovery mechanisms of the PE's transaction management part.

It is important to observe from Figure \ref{f:architecture} that, when H-Store executes a stored procedure, it requires a minimum of one round trip from the client to the partition engine, and one round trip from the partition engine to the execution engine if the stored procedure contains any SQL. In contrast, S-Store can trigger the next execution step within the appropriate layer using either {\em partition engine triggers (PE triggers)} or {\em execution engine triggers (EE triggers)} directly (see the red dashed horizontal arrows in Figure \ref{f:architecture}). This saves a fair amount of overhead, and, as we shall see later in Section \ref{sec:experiments}, it can even lead to a significant increase in overall transaction throughput.

In what follows, we describe the implementation of our architectural extensions in more detail.

\subsubsection{Streams} \label{sec:streams}

S-Store implements a stream as a time-varying, H-Store table. 
Using this approach, stream state is persistent and recoverable. Since tables represent unordered data sets, the order of tuples in a stream is captured based on tuple metadata (i.e., timestamps/tuple-id's and batch-id's). An atomic batch of tuples is appended to the stream table when it is placed on the corresponding stream, and conversely, an atomic batch of tuples is removed from the stream table as it is consumed by a downstream transaction in the workflow. The presence of an atomic batch of tuples within a stream can activate either a SQL plan fragment or a downstream transaction, depending on what ``triggers" are attached to the stream (described in Section \ref{sec:triggers}). In the case of the latter, the current stream table serves as input for the corresponding downstream transaction instance.

\subsubsection{Windows} \label{sec:windows}


Similarly to streams, windows are also implemented as time-varying, H-Store tables.
As in streams, arrival order for tuples in a window is tracked via tuple metadata.

As opposed to most traditional streaming systems that processed windows incrementally, in S-Store, SQL queries over windows operate one full window at a time. For a sliding window, a new full window becomes available every time that window has slide-worth of new tuples. Therefore, when new tuples are inserted into a window, they are flagged as ``staged" until slide conditions are met. Staged tuples are not visible to any queries on the window, but are maintained within the window.
Upon sliding, the oldest tuples within the window are removed, and the staged tuples are marked as active in their place. All window manipulation is done at the EE level, and output can be activated using an EE trigger (described in Section \ref{sec:triggers}).

Due to the invisible ``staging'' state of a window table as well as the transaction isolation rules discussed earlier in Section \ref{sec:isolation}, special scoping rules are enforced for window state. A window table is not allowed to be accessed in general by transaction executions other than those of the stored procedure that defined it. In fact, a window table must only be visible to future transaction executions of the stored procedure that contains it. As a consequence, one is not allowed to define PE triggers on window state, but only EE triggers. In other words, windows must be contained within the transaction executions of single stored procedures and must not be shared across other stored procedures in the workflow.

\subsubsection{Triggers} \label{sec:triggers}

Triggers are the fundamental construct in S-Store that enables push-based, data-driven processing needed by our streaming transactions as well as by workflows. A trigger is associated with a stream or a window table. When new tuples are appended to such a table, downstream processing will be automatically activated depending on where in the system stack the trigger is located. The alternative to triggers would be to poll for newly-arriving tuples, which would be highly inefficient.

There are two types of triggers in S-Store:

\begin{compactitem}

\item {\em Partition engine (PE) triggers} can only be attached to stream tables, and are used to activate downstream stored procedures upon the insertion of a new atomic batch of tuples on the corresponding streams. As the name implies, PE triggers exist to create a push-based workflow within the partition engine by eliminating the need to return back to the client to activate downstream stored procedures. In Figure \ref{f:architecture}, the horizontal arrows between stored procedures inside the PE layer denote PE triggers.

\item {\em Execution engine (EE) triggers} can be attached to stream or window tables, and are used to activate SQL queries within the execution engine. These triggers occur immediately upon the insertion of an atomic batch of tuples in the case of a stream, and upon the insertion of an atomic batch of tuples that also cause a window to slide in the case of a window. The SQL queries are executed within the same transaction instance as the batch insertion which triggered them, and can also activate further downstream EE triggers. EE triggers are designed to eliminate redundant communication between EE and PE layers. In Figure \ref{f:architecture}, the horizontal arrows between SQL queries inside the EE layer denote EE triggers.

\end{compactitem}

S-Store provides automatic garbage collection mechanisms for tuples that expire from stream or window state, after any triggers associated with them have all been fired and executed.

\subsubsection{Streaming Scheduler} \label{sec:scheduler}

Being an OLTP database that implements the traditional ACID model, the H-Store scheduler can execute transaction requests in any order. On a single H-Store partition, transactions run in a serial fashion by design \cite{hstore}. H-Store serves transaction requests from its clients in a FIFO manner by default.

As we discussed in Section \ref{sec:isolation}, streaming transactions and workflows require transaction executions for dependent stored procedures to be scheduled in an order that is consistent with the workflow graph (i.e., not necessarily FIFO). Additionally,
the application can specify (via defining nested transactions) additional isolation constraints, especially when shared table state among streaming transactions is involved.
The simplest solution to address these scheduling needs on a single partition is to adopt the most constrained approach in which the TE's in a workflow for a given input batch will always be executed in an order consistent with a topological ordering of the workflow. This will always ensure a correct, deterministic execution sequence. S-Store currently implements this approach by short-circuiting H-Store's FIFO scheduler in a way to fast-track stored procedures invoked via PE triggers to the front of the transaction request queue. This enables such TE's to be scheduled right away, preventing any other TE's from interleaving the workflow.

Although somewhat restrictive, we have found this approach to be practical in that it is amenable to a low-overhead implementation in H-Store and good enough to support all the S-Store use cases and benchmarks that we have studied (see Section \ref{sec:experiments}). As we are generalizing S-Store's implementation to the multi-node setting, we will also be extending its scheduler to consider other potential execution orders than FIFO and serial.

\subsubsection{Recovery Mechanisms} \label{sec:recoverymechanisms}

As described in Section \ref{sec:fault}, S-Store provides two different recovery options: (i) {\em strong recovery} is guaranteed to produce exactly the same state as was present before the failure, and (ii) {\em weak recovery} will produce a legal state that could have existed, but is not necessarily the exact state lost. Both of these options leverage periodic checkpointing and command-logging mechanisms of H-Store. However, they differ in terms of which transactions are recorded in the command-log during normal operation and how they are replayed during crash recovery. 

\vspace{0.05in}
\noindent
{\bf Strong Recovery.}
In strong recovery, as in H-Store, we log all transactions, both OLTP and streaming, in the command-log along with their input arguments. When a failure occurs, we replay the command-log in the order in which the transactions committed and were thus recorded in the command-log, starting from the latest snapshot. This will guarantee that the reads-from and the writes-to relationships between the transactions are strictly maintained. However, before this replay, we must first disable all PE-triggers so that the execution of a stored procedure does not redundantly trigger the execution of its successor(s) in the workflow. Once triggers are disabled, the snapshot is applied, and recovery from the command-log can begin. Because every transaction is logged in strong recovery, it is important that PE-triggers remain turned off to ensure that each interior transaction is only run once during the log replay. Failing to do this would create duplicate invocations, and thus potentially incorrect results.

When recovery is complete, we turn PE-triggers back on. At that point, we also check if there are any stream tables that contain tuples in them. For such streams, PE-triggers will be fired to activate their respective downstream transactions. Once those transactions have been queued, then the system can resume normal operation.

\vspace{0.05in}
\noindent
{\bf Weak Recovery.}
In weak recovery, the command-log need not record all stored procedure invocations, but only the ones that ingest streams from the outside (i.e., border stored procedures). We then use a technique similar to {\em upstream backup} \cite{hwang} to re-invoke other (i.e., interior) previously committed stored procedures. The basic idea behind upstream backup is to cache all the data at the inputs to a workflow so that in the event of a failure, the system can replay them in the same way that it did on first receiving them in the live system. Because the streaming stored procedures in an S-Store workflow have a well-defined ordering among them, the replay will necessarily create a correct execution history. While transactions may not be scheduled in the exact order that took place on the original run, some legal transaction order is ensured.

When recovering using weak recovery, we must first apply the snapshot, as usual. However, before applying the command log, S-Store must first check existing streams for data recovered by the snapshot, and fire any PE-triggers associated with those streams. This ensures that interior transactions that were run post-snapshot but not logged are re-executed. Once these triggers have been fired, S-Store can begin reading the log. Unlike for strong recovery, we do not need to turn off PE-triggers during weak recovery. In fact, we rely on PE-triggers for the recovery of all interior stored procedures, as these are not recorded in the command-log. This works fine as long as results are returned through committed tables.

Upstream backup is not a new technique. In fact, it was initially proposed as part of our Aurora/Borealis stream processing engine \cite{hwang}. However, we have found it easier to apply in a transactional setting. The fact that Aurora/Borealis did not have a transaction model made it difficult to know when it was safe to trim backup queues. In S-Store, however, we know that when a transaction commits, the input tuples associated with that transaction are no longer necessary. In other words, the existence of well-defined transactional units (atomic batches and stored procedures) simplifies the problem of queue trimming with upstream backup.

Weak recovery is a more light-weight alternative than strong recovery, since it need not log all committed transactions. Thus, it requires fewer writes to the command-log, at the expense of possibly longer recovery times. Section \ref{sec:recovery-exp} provides an experimental comparison of our strong and weak recovery mechanisms.

\section{Experiments} \label{sec:experiments}

\begin{figure}[t]
\begin{center}
\subfigure[EE Trigger Micro-benchmark]{
    \includegraphics[width=0.45\textwidth]{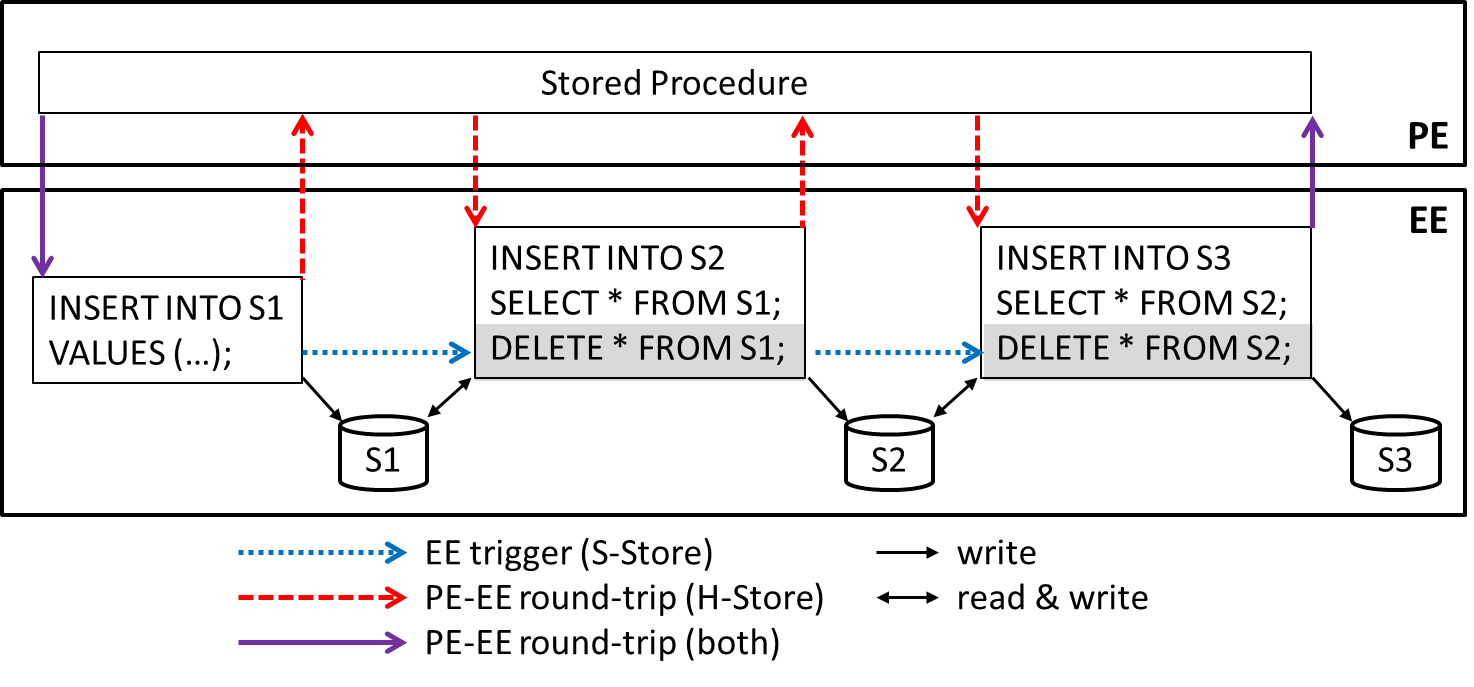}
    \label{f:EE-trigger-mb}
} \\
\subfigure[EE Trigger Result]{
    \includegraphics[width=0.45\textwidth]{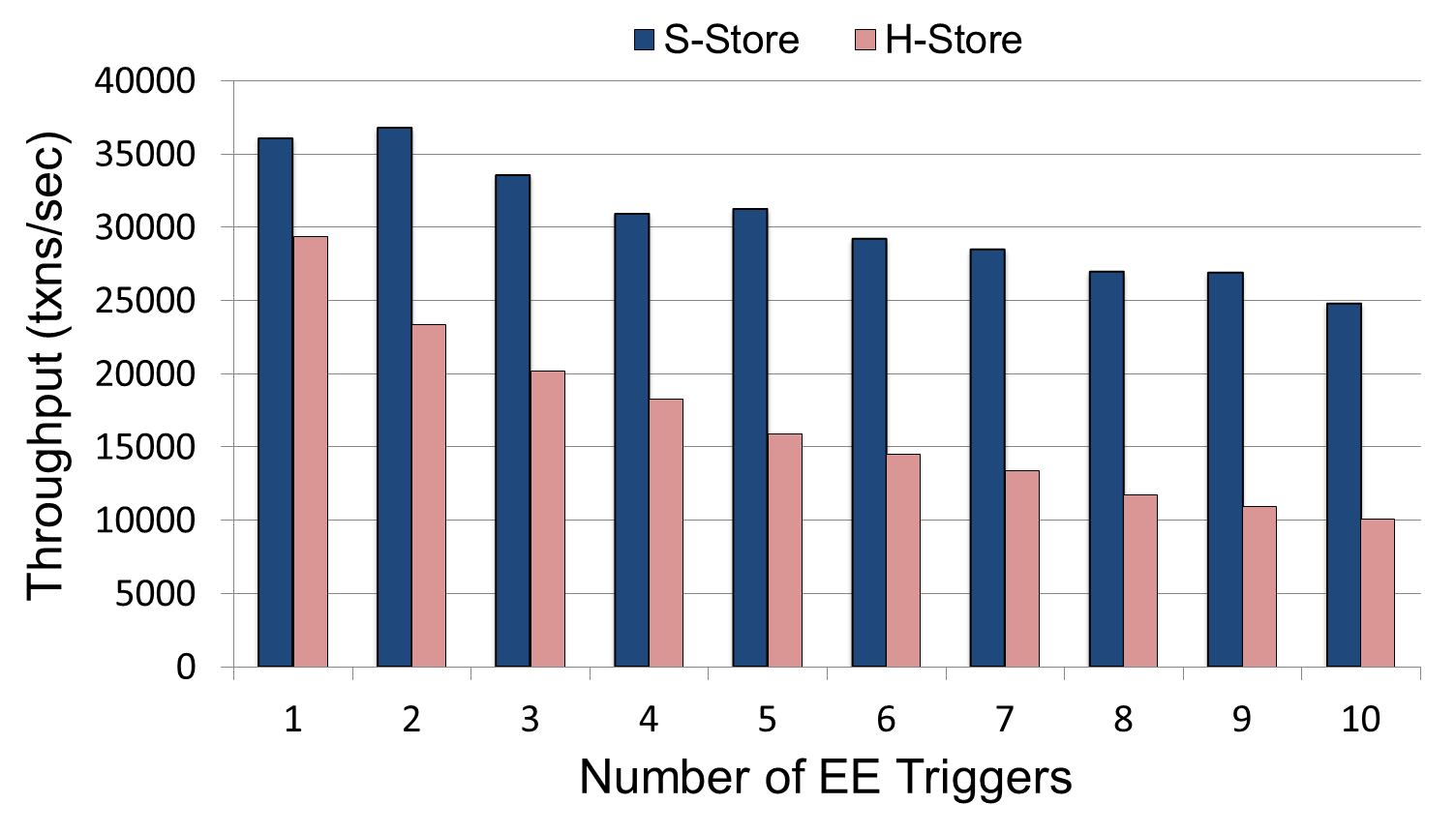}
    \label{f:EE-trigger-exp}
}
\caption{Execution Engine Triggers}
\label{f:EE-triggers}
\vspace{-0.3in}
\end{center}
\end{figure}

\begin{figure}[t]
\begin{center}
\subfigure[PE Trigger Micro-benchmark]{
    \includegraphics[width=0.41\textwidth]{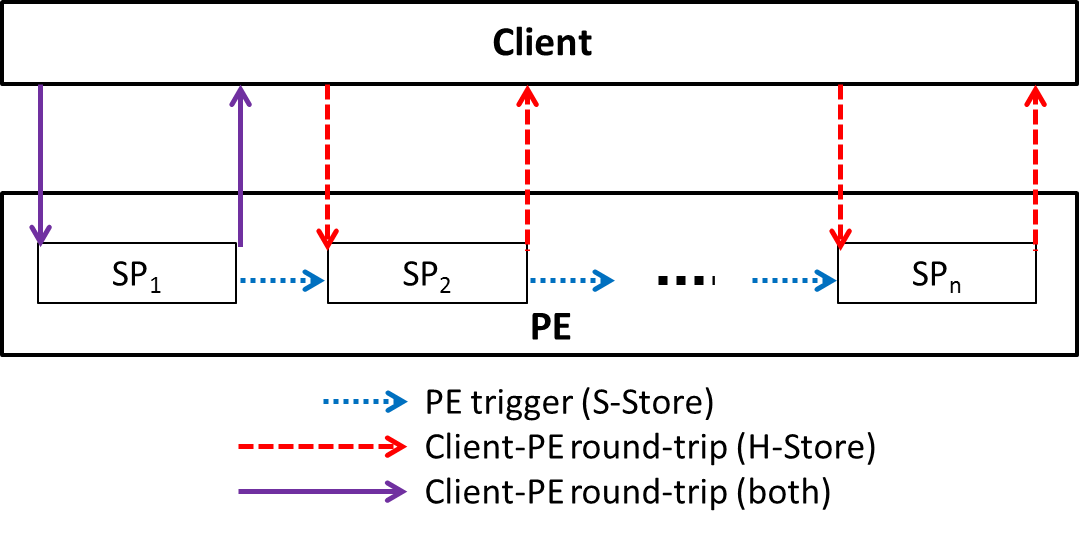}
    \label{f:PE-trigger-mb}
} \\
\subfigure[PE Trigger Result]{
    \includegraphics[width=0.45\textwidth]{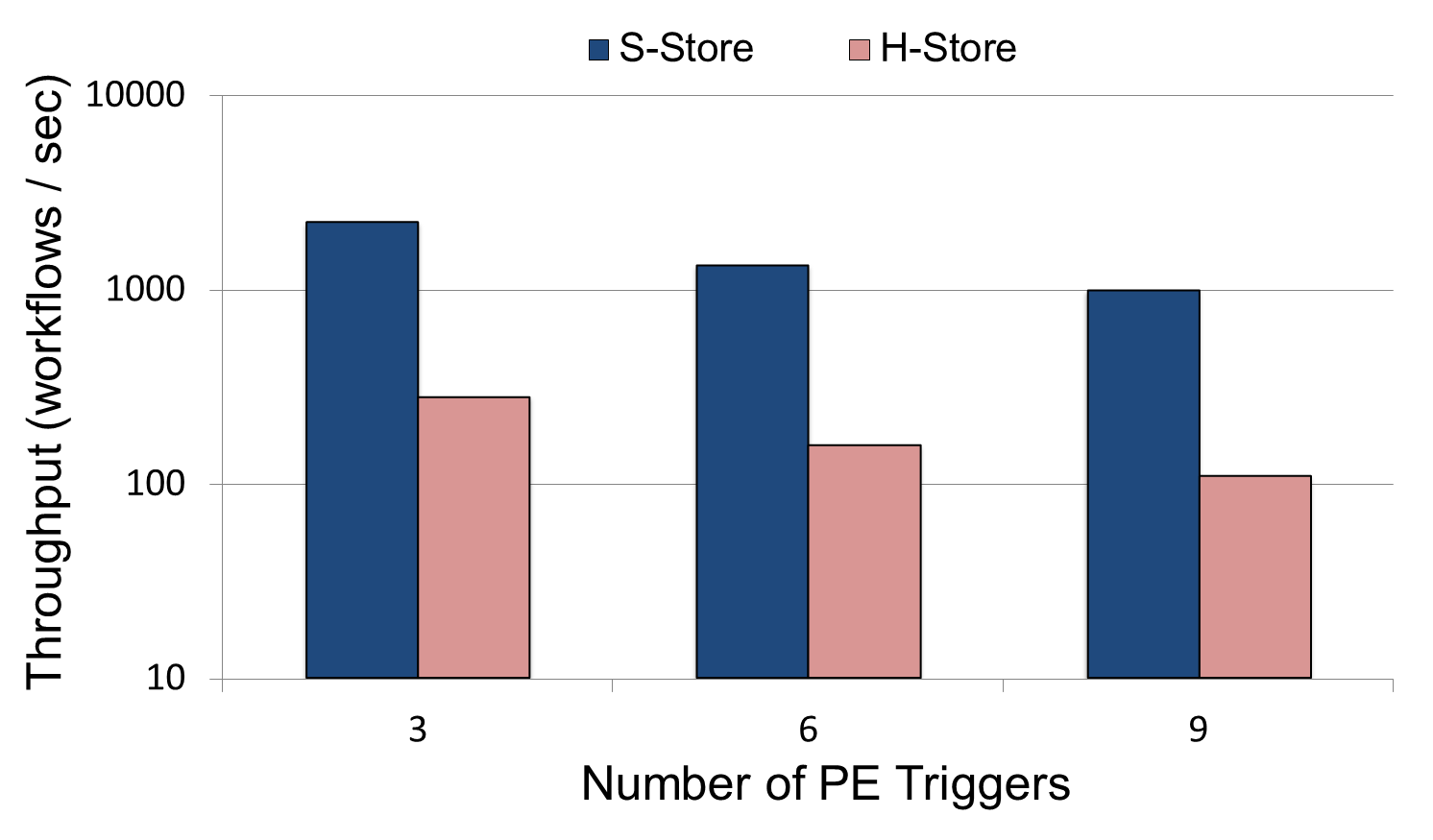}
    \label{f:PE-trigger-exp}
}
\caption{Partition Engine Triggers}
\label{f:PE-triggers}
\vspace{-0.3in}
\end{center}
\end{figure}

In this section, we present the results of our experimental study that evaluates S-Store with respect to existing  alternatives in OLTP and stream processing. More specifically, we compare S-Store to H-Store, Spark Streaming and Storm in terms of overall transaction throughput on a number of workloads. Sections \ref{sec:EE-triggers}-\ref{sec:recovery-exp} explore a number of micro-benchmarks that focus on evaluating specific architectural features of S-Store in comparison to its base system H-Store (e.g., triggers, windows, recovery modes).  The benchmarks described in Sections \ref{sec:voter}-\ref{sec:storm-spark} are based on our leaderboard maintenance application (described in Section \ref{sec:example}). This represents a more realistic use case, and we use it to compare S-Store to H-Store, Storm and Spark Streaming. The benchmark details are described below together with their corresponding results. We also provide early insight into S-Store's scalability in Section \ref{sec:multisite}.  Here we implement a subset of the Linear Road benchmark \cite{arasu2004linear} in order to demonstrate S-Store's ability to effectively parallelize streaming workflows in the case of highly partitionable workloads.

To properly evaluate streaming workloads, we record throughput in terms of ''workflows per second''.  This number represents the number of full workflows that complete end-to-end, regardless of the number of transactions required to complete those workflows.  In cases where workflows necessarily consist of only a single transaction, we simplify our unit to ''transactions per second''. All experiments were run on a cluster of Intel Xeon machines, each containing 64 cores and 264 GB of memory. Because we focus on single-node S-Store deployments in this paper and due to the partitioned architecture of S-Store, effectively only a single core is used for data access (with the exception of Section \ref{sec:multisite}). The experiments were run using a single non-blocking client which asynchronously sends requests to the system. Logging was disabled unless otherwise specified.

\subsection{Execution Engine Triggers} \label{sec:EE-triggers}

In this experiment, we evaluate the benefit of S-Store's EE triggers. The micro-benchmark contains a single stored procedure that consists of a sequence of SQL statements that simulate an input stream flowing through a set of query stages (Figure \ref{f:EE-trigger-mb}). In S-Store, these SQL statements can be activated using EE triggers such that all execution takes place inside the EE layer. H-Store, on the other hand,
must submit the set of SQL statements (an insert and a delete) for each query stage as a separate execution batch from PE to EE. Figure \ref{f:EE-trigger-mb} illustrates the case for 3 streams and query stages. S-Store's EE triggers enable it to trade off trigger execution cost for reducing the number of PE-to-EE round-trips (e.g., 2 triggers instead of 2 additional round-trips). Note also that the deletion statements are not needed in S-Store, since garbage collection on streams is done automatically as part of our EE trigger implementation.

Figure \ref{f:EE-trigger-exp} shows how transaction throughput changes with varying number of EE triggers. S-Store outperforms H-Store in all cases and its relative performance further increases with increasing number of EE triggers, reaching up to a factor of 2.5x for 10 triggers.


\subsection{Partition Engine Triggers} \label{sec:PE-triggers}

This experiment compares the performance of S-Store's PE triggers to an equivalent implementation in H-Store, which has no such trigger support in its PE. As illustrated in Figure \ref{f:PE-trigger-mb}, the micro-benchmark consists of a workflow with a number of identical stored procedures that must execute one after the other. We assume that the workflow must execute in exact sequential order, as is the case with many streaming workflows that require the maintenance of consistent state across the workflow. In H-Store, the scheduling request of a new transaction must come from the client, and because the workflow order of these transactions must be maintained, transactions cannot be submitted asynchronously. Serializing transaction requests severely limits H-Store's performance, as the engine will be unable to perform meaningful work while it waits for a client request. In S-Store, a PE trigger can activate the next transaction directly within the PE and can prioritize these triggered transactions ahead of the current scheduling queue using its streaming scheduler. Thus, S-Store again trades off additional trigger execution overhead for avoiding unnecessary round-trips to the client layer.

Figure \ref{f:PE-trigger-exp} shows how transaction throughput (measured in workflows/sec and plotted in log-scale) changes with increasing workflow size (shown in number of PE triggers for S-Store). 
As can be seen, H-Store's throughput tapers early due to the PE's need to wait for the client to determine which transaction is next to schedule.  
S-Store is able to process roughly an order of magnitude more workflow instances per second thanks to its PE triggers. This performance benefit becomes even more pronounced as the number of transactions per workflow increases.

\vspace{2mm}
\subsection{Windows} \label{sec:windows-exp}

\begin{figure}[!t]
\begin{center}
\subfigure[b][Window Micro-benchmarks]{
   \includegraphics[width=0.45\textwidth]{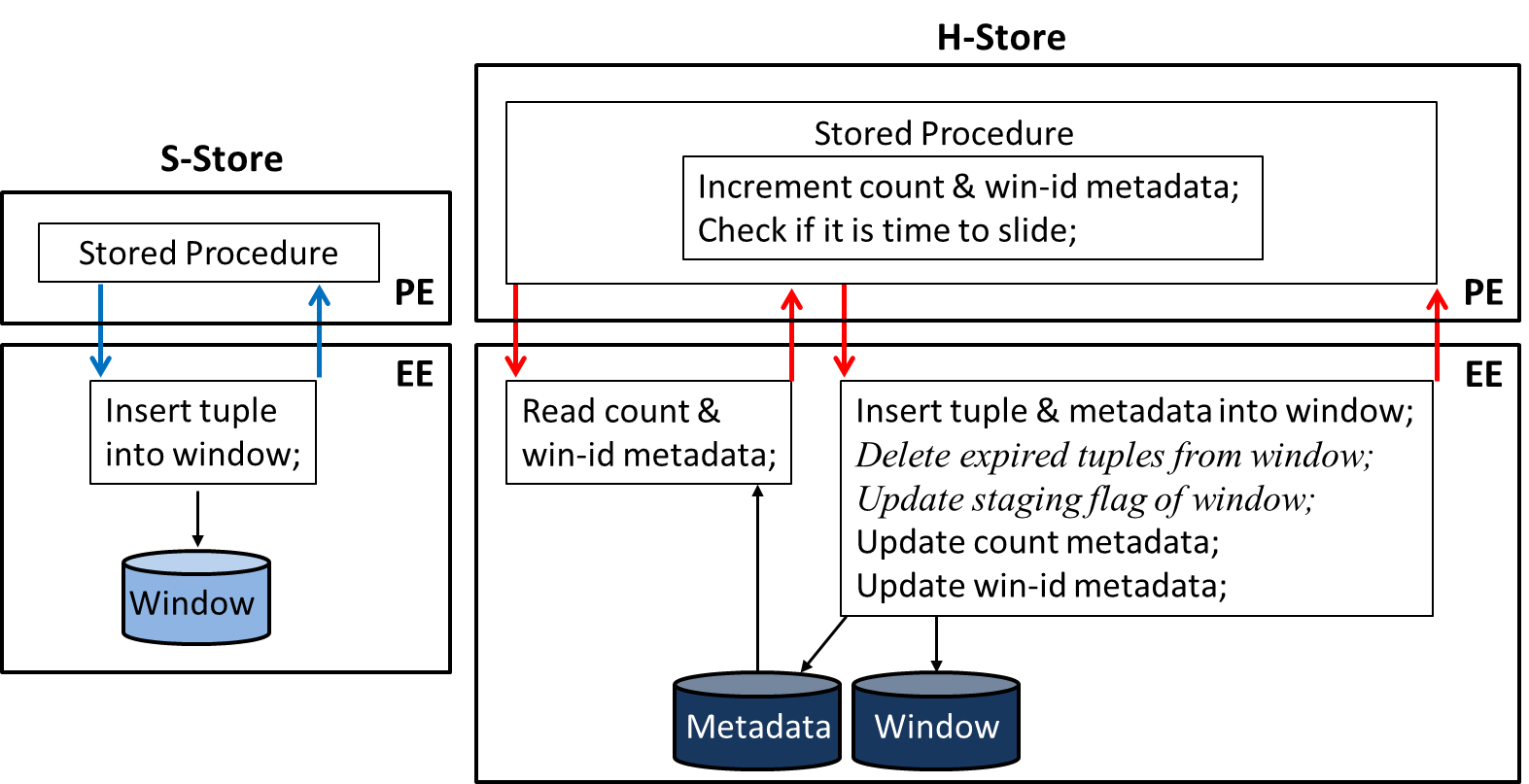}
    \label{f:window-mb}
} \\
\vspace{-0.1in}
\subfigure[b][Window Result]{
   \includegraphics[width=0.5\textwidth]{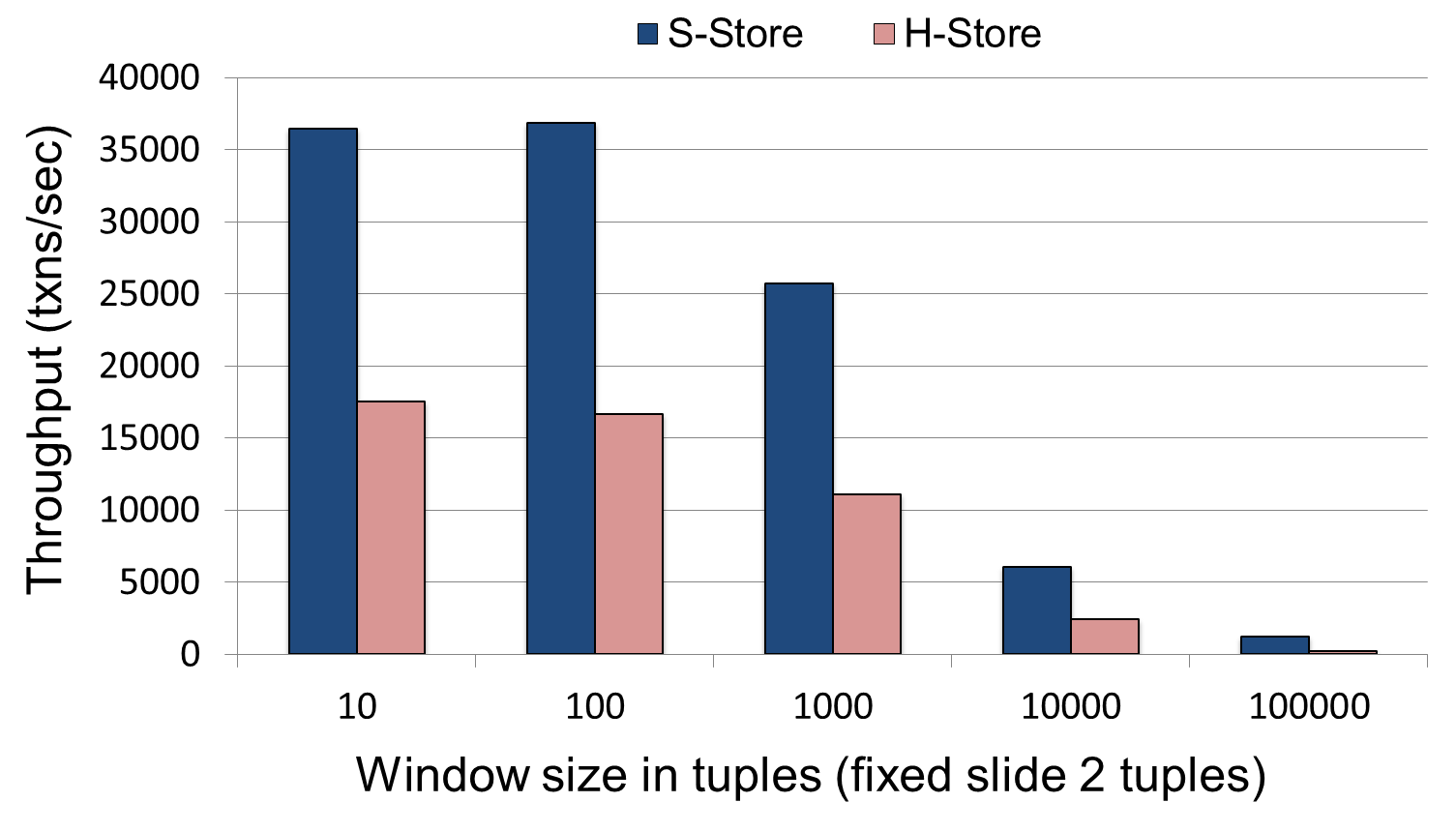}
    \label{f:window-size-exp}
} \\
\vspace{-0.1in}
\caption{Windows}
\vspace{-0.3in}
\label{f:windows}
\end{center}
\end{figure}

In this micro-benchmark, we compare native windowing support provided by S-Store within its EE to a comparable H-Store implementation without such built-in windowing support. As mentioned in Section \ref{sec:windows}, S-Store employs a tuple-wise ordering strategy and flags new tuples for ''staging'' as they arrive, keeping them inactive until the window slides.  The most fair strategy for H-Store is to use a similar technique, in which separate columns are maintained to indicate the absolute ordering of tuples, as well as a flag denoting which tuples are active within the window. If enough tuples have arrived to slide the window, then ''staged'' tuple flags are removed and expired tuples are deleted.

Figure \ref{f:window-mb} depicts the micro-benchmark we used in this experiment, which simply inserts new tuples into a tuple-based sliding window and maintains the window as tuples continue to arrive. As seen on the left-hand side of the figure, S-Store can implement this micro-benchmark simply by defining the window with its desired parameters and issuing insertion queries over it. An equivalent implementation in H-Store (shown on the right-hand side of Figure \ref{f:window-mb}) would require the maintenance of both a window and a metadata table, with a two-staged stored procedure to manage the window state using a combination of SQL queries and Java logic.

Figure \ref{f:window-size-exp} shows the transaction throughput achieved by S-Store and H-Store as window size varies. Our native windowing implementation grants us roughly a 2x throughput gain over the naive implementation of H-Store. This gain primarily comes from maintaining window statistics (total window size, number of tuples in staging, etc.) within the table metadata, allowing window slides and tuple expirations to happen automatically when necessary. It is also important to note that variations in window size provide a much larger contribution to the performance difference than window slide variations. This is because slide frequency only affects the frequency of two SQL queries associated with window management (shown in italic in Figure \ref{f:window-mb}).


\subsection{Recovery Mechanisms} \label{sec:recovery-exp}

As described earlier in Sections \ref{sec:fault} and \ref{sec:recoverymechanisms}, S-Store provides two methods of recovery: {\em strong recovery}, in which all transactions are individually written to the command-log, and {\em weak recovery}, a version of upstream backup in which only border stored procedures are logged, and PE triggers allow interior stored procedures to be automatically activated during log replay. We now investigate the performance differences caused by logging overhead and recovery-time for these two methods.

\begin{figure*}[t]
\begin{center}
\subfigure[Logging]{
\includegraphics[width=0.45\textwidth]{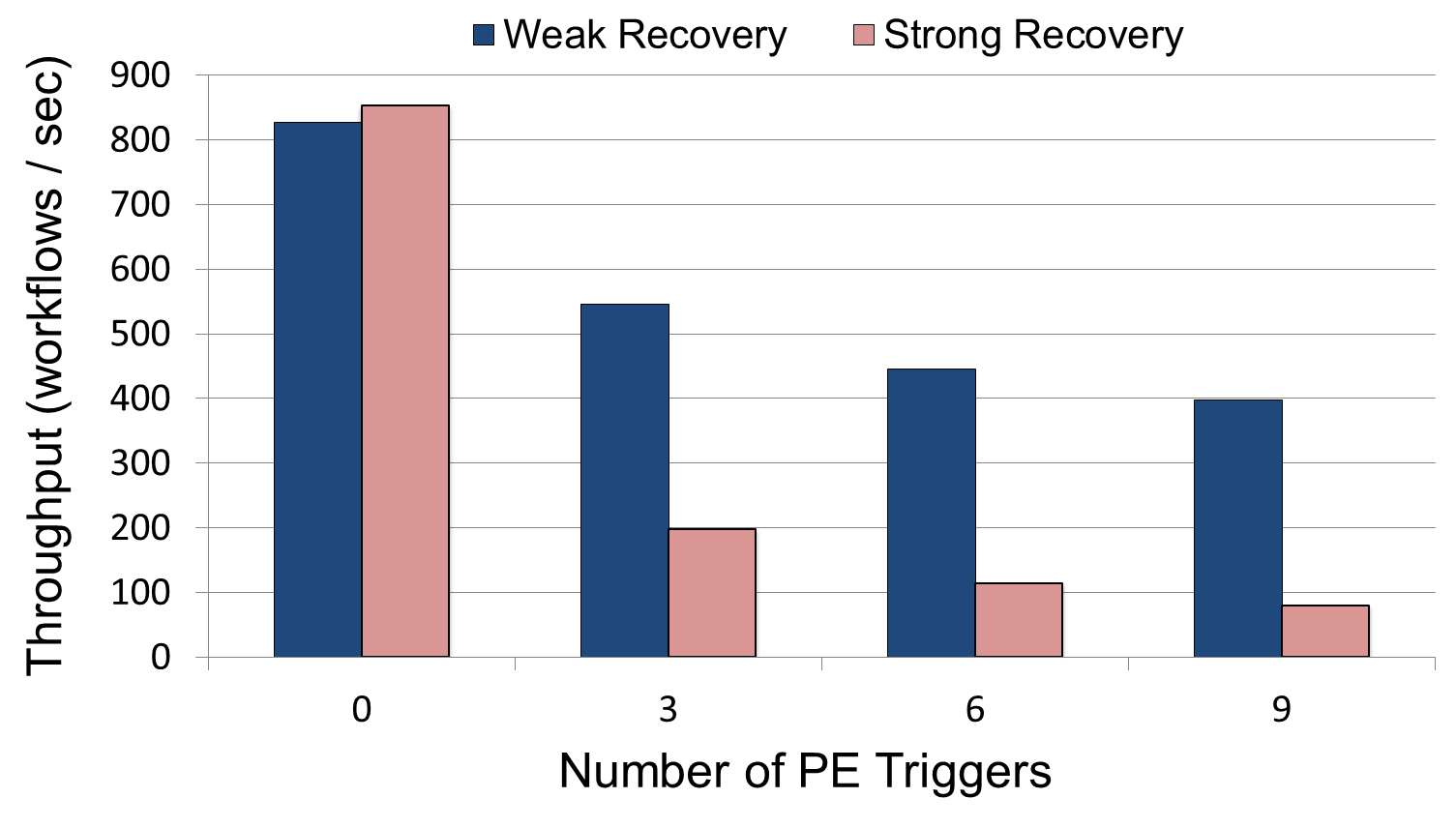}
\label{f:logginggraph}
}
\subfigure[Recovery]{
\includegraphics[width=0.45\textwidth]{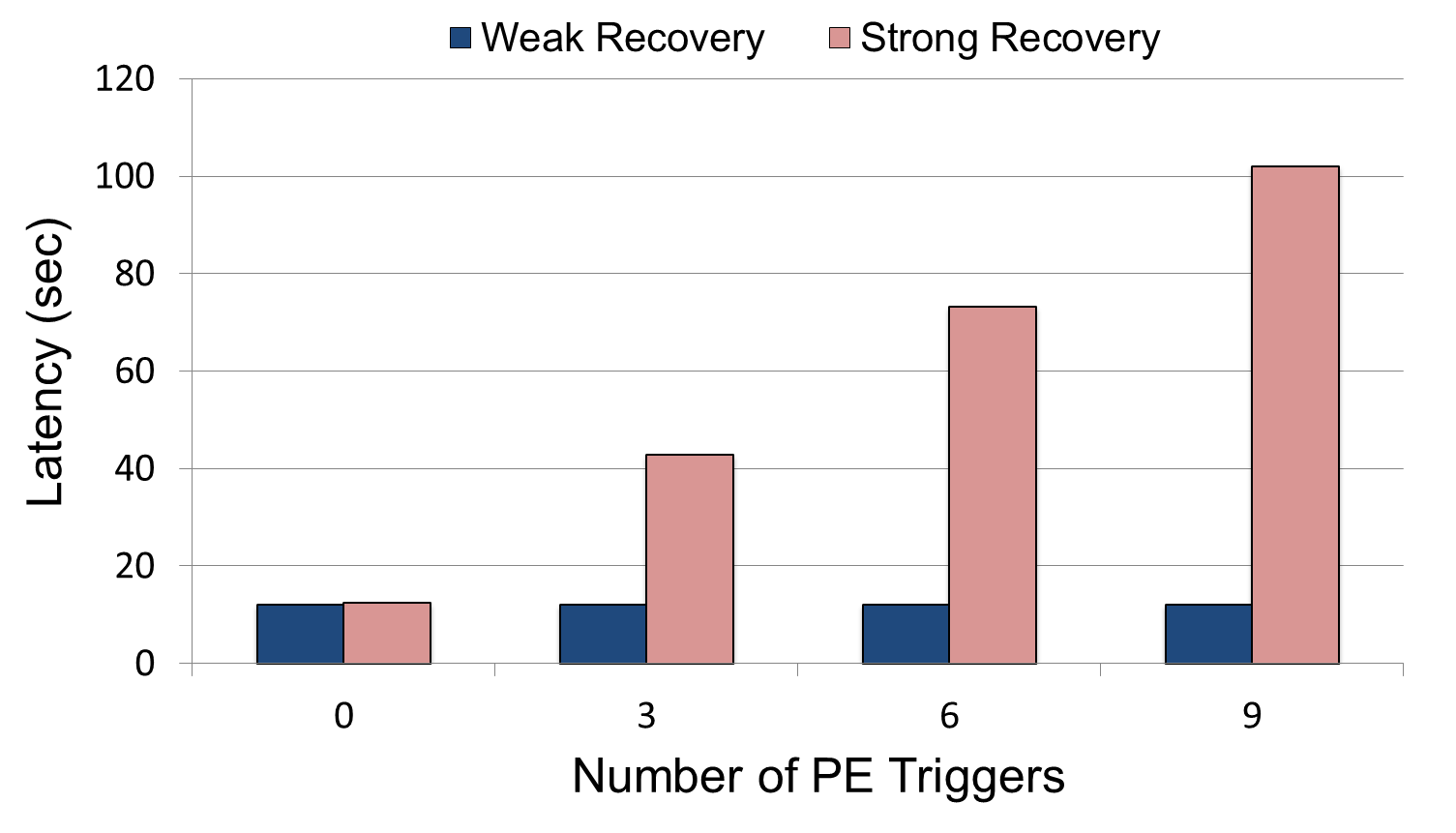}
\label{f:recoverygraph}
}
\vspace{-0.3in}
\caption{Recovery Mechanisms}
\label{f:logging-recovery}
\vspace{-0.2in}
\end{center}
\end{figure*}

For this experiment, we use the same micro-benchmark presented in Section \ref{sec:PE-triggers} (Figure \ref{f:PE-trigger-mb}), using a variable number of PE triggers within the sequence. Ordinarily in H-Store, higher performance is achieved during logging by grouping many transactions and committing them together, writing their log records to disk in batches. There are situations, however, when group committing is not possible, and each transaction must be logged and committed separately.  For trigger-heavy workloads, weak recovery can accomplish a similar run-time effect even without the use of group commit. As seen in Figure \ref{f:logginggraph}, without group commit, logging quickly becomes a bottleneck in the strong recovery case.  Each transaction is logged, so the workflow throughput quickly decreases as the number of transactions per workflow increases.  By contrast, weak recovery writes to the log for only the border transactions, allowing up to 4x the throughput as it writes a smaller fraction of log records to disk.

Figure \ref{f:recoverygraph} shows the striking result that weak recovery not only achieves better throughput during normal operation, but it also provides lower recovery time. Typically during recovery, the log is read by the client and transactions are submitted sequentially to the engine. Each transaction must be confirmed as committed before the next can be sent. Because weak recovery activates non-border transactions within the engine, the transactions can be confirmed without a round-trip to the client. As a result, latency stays roughly constant for weak recovery even for workloads with large numbers of PE triggers, while for strong recovery it increases linearly with the number of PE triggers.

\subsection{Leaderboard Maintenance} \label{sec:voter}

\begin{figure}[t]
\centering
\includegraphics[width=0.45\textwidth]{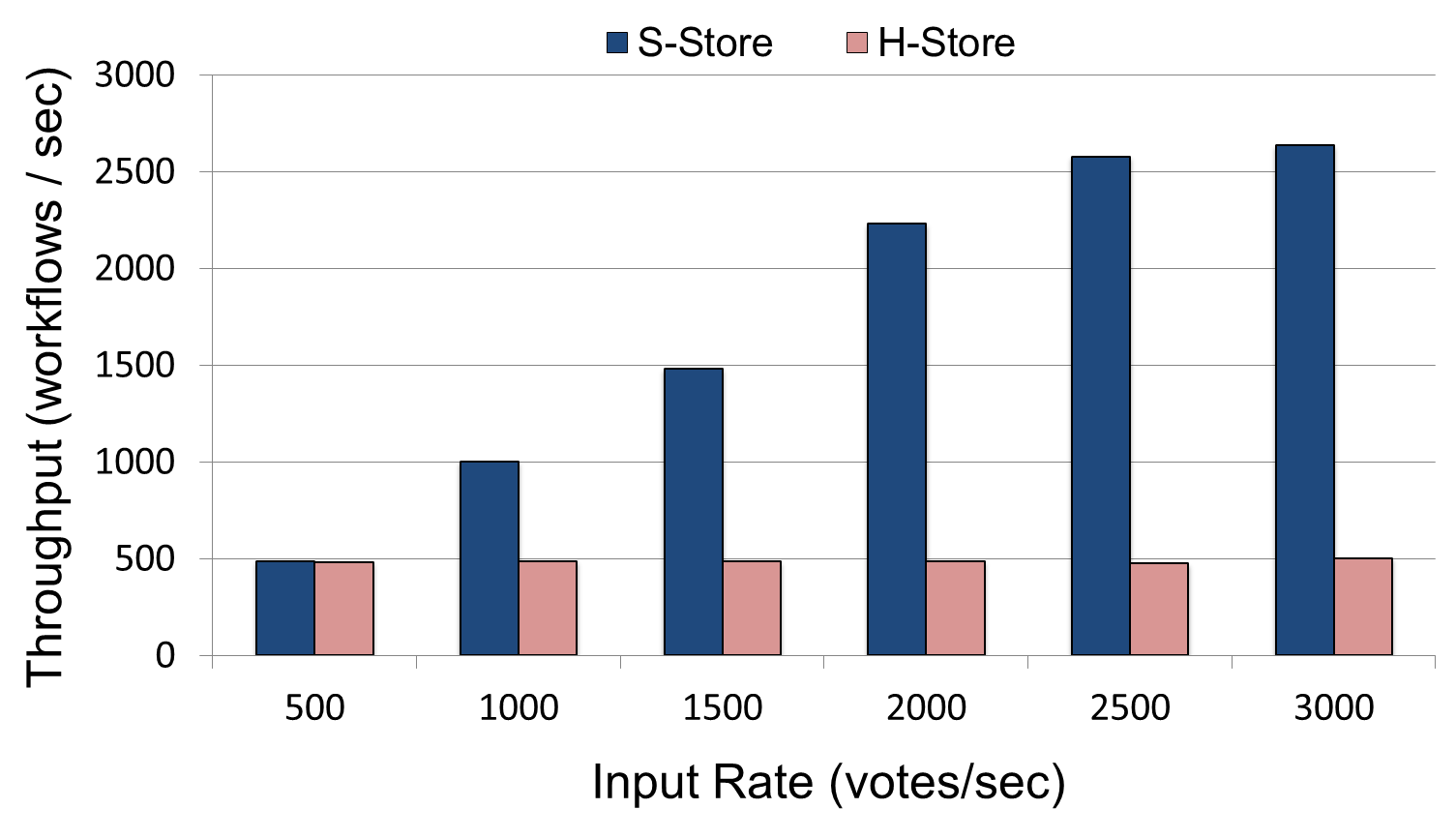}
\vspace{-0.1in}
\caption{Leaderboard Maintenance}
\vspace{-0.2in}
\label{f:votergraph}
\end{figure}

This experiment compares S-Store's throughput performance aga\-inst that of H-Store on our leaderboard maintenance example previously described in Section \ref{sec:intro}. As was shown in Figure \ref{f:LB-workflow}, the workflow contains three separate stored procedures: one to validate and insert a new vote, a second to maintain the leaderboard, and a third to delete a candidate if necessary. In order to ensure the correctness of the result in the presence of shared tables, as well as to maintain consistency of the tables across the workflow, these three stored procedures must execute in sequence for each new vote. This benchmark practices all of the architectural additions of S-Store described in Section \ref{sec:implementation}.

As seen in Figure \ref{f:votergraph}, much like the results of our micro-benchmarks, S-Store maintains a significant throughput advantage over its OLTP predecessor. More specifically, as input rate in terms of votes/sec increases, both systems begin by displaying the same workflow throughput. However, as the rate goes beyond 500 workflows per second, H-Store is unable to keep up, while S-Store continues to successfully process nearly 3000 votes/sec. As expected, the main reason for this result is S-Store's use of PE triggers, which avoids redundant round-trips to the client. The streaming scheduler also ensures that transactions are run in proper sequence, while also allowing future transactions to be asynchronously queued. H-Store has no method of providing this queueing guarantee.  Thus, to prevent incorrect results, it must wait for the client to explicitly determine the next transaction to queue based on the result of the previous transaction. EE triggers and native windowing also contribute to the throughput increase, though to a lesser degree than PE triggers and the streaming scheduler.

\subsection{Modern Streaming Systems Comparison} \label{sec:storm-spark}

In order to compare S-Store against state-of-the-art streaming systems, we chose to implement two variants of our leaderboard maintenance benchmark with Storm \cite{storm} and Spark Streaming \cite{zaharia13}. For the Storm implementation, we opted to use Trident\cite{trident-wiki}, an extension of Storm that supports stateful stream processing and exactly-once semantics. The following experiments compare the single-node performance of S-Store against Storm with Trident and Spark Streaming.

\subsubsection{Spark Streaming} \label{sec:spark}
As Spark Streaming does not include any notion of transactions and supports only time-based windows, we have simplified our benchmark to a single stored procedure which handles vote validation and recording while maintaining a time-windowed leaderboard (10 second windows, sliding every 1 second). The Spark benchmark uses Redis for vote state and RDDs for durability.

Like most streaming systems, Spark Streaming inherently has no concept of a true ACID transaction; we consider a Spark batch to be the proper analog to a transaction within S-Store. In our S-Store experiments, each vote is submitted as its own separate transaction (i.e., input batch size is a single tuple). We considered submitting only a single vote per Spark batch and minimizing the duration per batch in order to gain a similar level of isolation and atomicity to S-Store transactions. However, it quickly became apparent that Spark's performance would be extremely poor under these conditions. Spark Streaming primarily gains its performance through batch-processing and parallelization, so to remove both elements would provide an unfair comparison. 

\subsubsection{Storm with Trident} \label{sec:storm}
Ordinarily, Storm is stateless and guarantees either at-least-once or at-most-once semantics. Trident improves on Storm by offering much stronger transactional guarantees, adding stateful, incremental processing on top of any database \cite{trident-wiki}, as well as exactly-once processing. In Trident, we create two "Bolts" (Storm's equivalent of stored procedures): one that validates votes and another that maintains a leaderboard. As Trident has no notion of windows, sliding window functionality must be implemented with temporal state management and event triggering. State management for vote validation requires indexing; this is built using Memcached. Additionally, workflows are logged asynchronously using Storm's logging capabilities to ensure durability of processed streaming data. 

\subsubsection{Comparison} \label{sec:sstore-comparison}
\begin{figure}[t]
\centering
\includegraphics[width=\columnwidth]{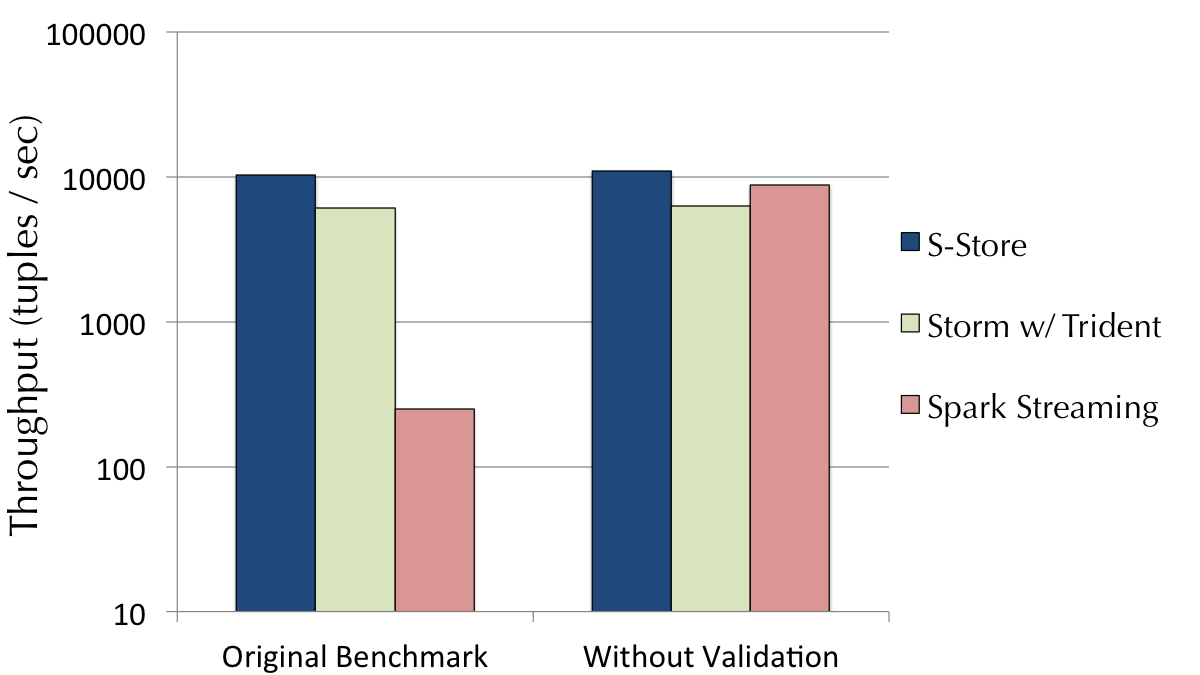}
\vspace*{-12pt}
\caption{Voter w/ Leaderboard on Modern SDMSs}
\vspace{-0.2in}
\label{f:sparkgraph}
\end{figure}

For S-Store, we ran experiments on the original transactional version of the benchmark (one vote per batch) with logging. As shown in the left-hand side of Figure \ref{f:sparkgraph}, S-Store and Trident perform comparably, but the state handling and interface that provides exactly-once guarantees impairs Storm's performance \cite{storm}. Spark exhibits surprisingly lower performance on this particular workload against S-Store. This is mainly caused by the validation process, in which a new tuple's phone number is verified as having not previously voted. This operation requires comparison to all previous votes that have been recorded, and happens each time a new vote arrives. Spark Streaming provides no method of indexing over state, and thus validation becomes extremely expensive as the total number of votes grows. S-Store, on the other hand, is able to handle this task easily via an index on the ''phone number'' column of the Votes table, providing a lookup rather than a table scan.


In order to better compare against Spark's strengths, we created a second variant of the same benchmark that does not rely so heavily on indexed tables. This version is identical to the first, except that the vote validation against previous records is removed. The remaining portions of the workload are easily expressed as map-reduce operations (i.e., natural to express in Spark's programming model). As shown on the right-hand side of Figure \ref{f:sparkgraph}, neither S-Store nor Trident's performance show significant increase in throughput, proving that leaderboard maintenance remains the bottleneck for their performance. In the case of Trident, the lack of built-in windowing functionality curbs its overall performance. Spark Streaming performs over an order of magnitude better on this simplified workload than it does on the original. However, even after removing Spark's major bottleneck from the workload, S-Store is able to match both Spark and Storm's performance while providing much stronger transactional guarantees than either.


\subsection{Multi-Core Scalability} \label{sec:multisite}

\begin{figure}[t]
\centering
\includegraphics[width=0.9\columnwidth]{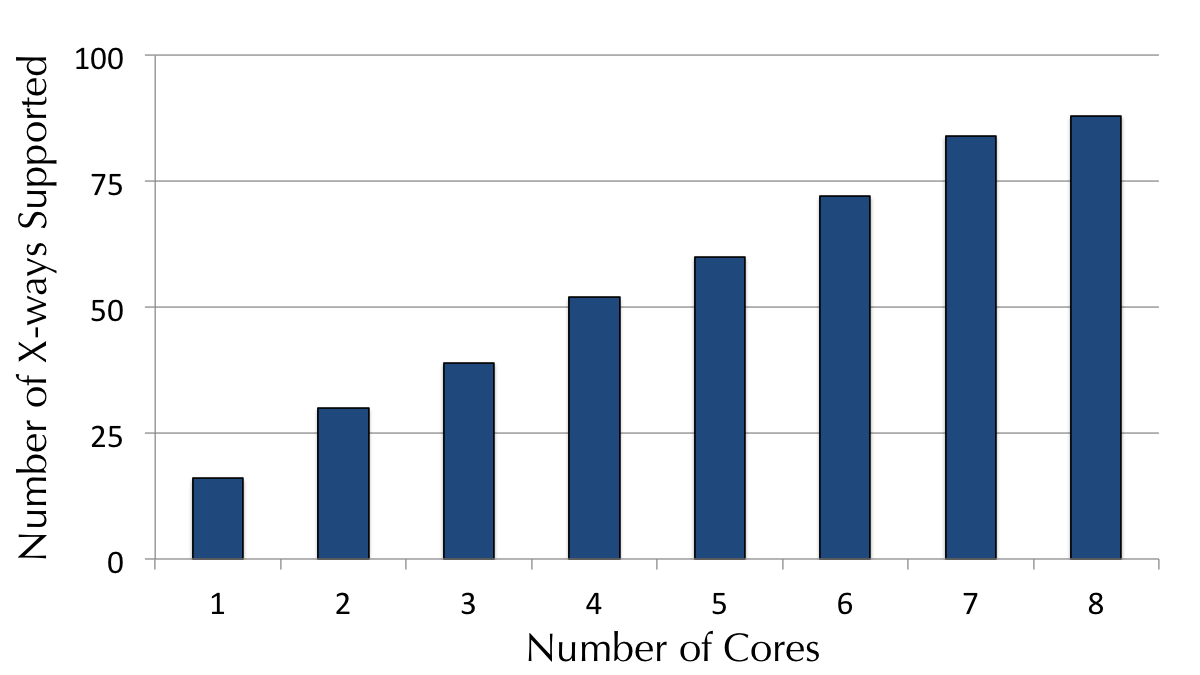}
\vspace*{-12pt}
\caption{Multi-core Scalability on S-Store}
\vspace{-0.2in}
\label{f:multicore}
\end{figure}

We are currently extending S-Store to operate on multiple nodes. This will allow us to scale to both multiple cores and machines by intelligently dividing the workload (both data and workflows) to be processed in parallel. While a complete solution is still in progress, S-Store is already able to leverage H-Store's distributed architecture described in Section \ref{sec:hstore} to parallelize workflow processing for highly partitionable workloads. Specifically, S-Store is able to partition an input stream onto multiple cores.  Each core runs TE's of the complete workflow in a serial, single-sited fashion for the input stream partition to which it is assigned. Our serial scheduling as well as both weak and strong recovery mechanisms apply to each core as described in Section \ref{sec:arch}.

In order to test S-Store's scalability on multiple cores for this type of workload scenario, we have implemented a subset of the Linear Road data stream benchmark \cite{arasu2004linear}. Linear Road simulates traffic on a specified number of "x-ways" and requires that notifications about tolls and accidents be dynamically delivered to vehicles based on their position reports. For the purposes of demonstrating partitioned stream workflows, it was sufficient to focus only on the queries over streaming position reports; thus, we excluded the historical queries of the benchmark. We simulated traffic data for 30 minutes, and evaluated performance by comparing the number of x-ways we are able to support while ensuring that all position reports are processed under a given latency threshold. Because we are running an abbreviated version of the benchmark, we set our latency threshold to be 1 second rather than the recommended 5 seconds.

We implemented this workload as a workflow that consists of two stored procedures. SP1 updates a vehicle's position and detects whether it has entered a new segment. If it has, SP1 sends a notification to the vehicle of any tolls or accidents recorded for upcoming road segments in the 1st minute, and charges the vehicle tolls from the previous segment. SP1 is also responsible for detecting stopped cars and accidents. At the beginning of each minute, SP1 triggers SP2, which calculates the tolls for the previous minute and records the current statistics for every x-way table into a historical statistics table. SP2 also checks all accidents to see whether the vehicles involved have been removed from the scene. Accidents and tolls are all calculated relative to the events happening on a single x-way, and thus we distribute the x-ways evenly across partitions. Because each partition runs transactions serially, we expect to achieve roughly linear scalability as we increase the number of cores.

Figure \ref{f:multicore} confirms our expectations.  S-Store is able to support 16 x-ways per core when a single partition is used, and scales roughly linearly, with about 5-10 percent drop-off per added core due to additional partition maintenance.  At the upper end of the graph (8 cores), 11 x-ways per core are able to be supported.  S-Store is best able to support a number of x-ways that is divisible by the number of cores that are available.  For instance, 3 cores are able to support 39 x-ways, but adding a 40th will disproportionately increase latency as the load between cores becomes asymmetric.

\section{Related Work} \label{sec:related}


In the early 2000's, there was a lot of interest in the database community for stream processing. The main goal of this work was to process continuous queries with low latency as data streamed into the system. This was largely inspired by the emergence of sensor-based applications. Many academic prototypes (Aurora / Bore\-alis \cite{aurora, borealis}, STRE\-AM \cite{stream}, TelegraphCQ \cite{telegraphcq}, NiagraCQ \cite{niagracq}) were built, and several commercial products were spawned as a result of this work (e.g., TIBCO StreamBase, CISCO Truviso, SAP Coral8, IBM InfoSphere Streams, Microsoft StreamInsight, Oracle CEP). With the exception of STREAM and Coral8, these systems did not support an explicit notion of transactions. STREAM did not directly claim to have transactions, but its execution model was based on logical timestamps which could be interpreted as transaction IDs. Batches of tuples with the same timestamp were executed atomically. While this could be used to provide isolation, recovery was not discussed. Furthermore, modeling transactions as the execution of an entire workflow (query) did not allow finer-grained transaction definitions. Similarly, Coral8 provided so-called ''atomic bundles'' as a configurable isolation/recovery unit embedded in its execution model, but did not provide any transactional guarantees beyond "at least once" for processing events through the entire workflow. Furthermore, none of these early systems considered integrating stream processing with traditional OLTP-style query processing.


Fault tolerance issues have been investigated as stream processing systems have been moved into distributed settings \cite{borealis, telegraphcq}. A few fundamental models and algorithms have been established by this work \cite{hwang, balazinska, shah}, including the upstream backup technique that we leverage in our weak recovery mechanism.

There have also been several efforts in addressing specific transactional issues that arise in stream processing settings. For example, Golab et al. have studied the concurrency control problem that arises when a sliding window is advanced (write operation) while it is being accessed by a query (read operation) \cite{golab-edbt06}. This work proposes sub-windows to be used as atomic access units and two new isolation levels that are stronger than conflict serializability. Such a problem never arises in S-Store, since window state is accessed by a single TE at a time (and never by TEs of different SPs). As another example, Wang et al. have considered concurrency issues that arise when adding active rule support to CEP engines in order to monitor and react to streaming outputs \cite{wang-vldb11}. In this case, the rules may require accessing state shared with other queries or rules. This work defines a stream transaction as a sequence of system state changes that are triggered by a single input event, and proposes a timestamp-based notion of correctness enforced through appropriate scheduling algorithms. S-Store investigates transactional stream processing in a more general context than active CEP engines.

Botan et al.'s work was the first to recognize the need for an explicit transaction model to support queries across both streaming and stored data sources \cite{botan-edbt12}. This work proposed to extend the traditional page model \cite{weikum-book} to include streams of events (as time-varying relations) and continuous queries (as a series of one-time queries activated by event arrivals). As a result, each one-time query execution corresponds to a sequence of read/write operations, and operations from one or more such sequences can be grouped into transaction units as required by the application semantics. Transactions must then be executed in a way to ensure conflict serializability and event arrival ordering. Thus, this work focused on the correct ordering of individual read/write operations for a single continuous query, and not so much on transaction-level ordering for complex workflows like we do. As such, S-Store proposes a more general transaction model.



Recently, a new breed of stream processors has emerged. Unlike the majority of the earlier-generation systems, these do not adopt a select-project-join operator environment. Instead, they expect the user to supply their own operators (UDF's), and the system controls their execution in a scalable fashion over a cluster of compute nodes. Typically, these systems provide fault tolerance and recoverability, but do not support fully-ACID transactions. Representatives include Storm \cite{storm} and Spark Streaming \cite{zaharia13}.
  
Storm has been originally built to support Twitter's real-time computation needs. Like our workflows, computations are represented as {\em topologies}, i.e., directed graphs of operators called {\em bolts}.
Storm provides two types of semantic guarantees: {\em at least once} and {\em at most once}. In case of {\em at least once} semantics, each tuple is assigned a unique message-id and its lineage is tracked. For each output tuple $t$ that is successfully delivered by a topology, a backflow mechanism is used to acknowledge the tasks that contributed to $t$ with the help of a dedicated {\em acker bolt} added to the corresponding topology. The data source must hold the tuple until a positive ack is received and the tuple can be removed (similar to upstream backup \cite{hwang}). If an ack is not received within a given timeout period, then the source will replay the tuple again. Storm can only provide the weaker {\em at most once} semantics when the acknowledgment mechanism is disabled. Trident provides a higher-level programming abstraction over Storm which provides a stronger, {\em exactly once} processing guarantee based on automatic replication \cite{trident-wiki}. While these guarantees ensure some level of consistency against failures, they are not sufficient to support atomicity and isolation as in the case of ACID guarantees. Furthermore, Storm focuses on purely streaming topologies and thus lacks support for dealing with persistent state and OLTP transactions.

Spark Streaming extends the Spark batch processing engine with a new stream processing model called discretized streams (D-Streams) \cite{zaharia13}. The basic idea is to divide (mostly analytical) computations into a series of stateless, deterministic transformations over small batches of input tuples based on their arrival time intervals. Somewhat similar to STREAM, tuples are processed atomically within the time interval that they arrive in. All state in Spark Streaming is stored in in-memory data structures called Resilient Distributed Datasets (RDDs). RDDs have two key properties: they are immutable and partitioned. Immutability enables efficient asynchronous checkpointing as well as maintaining different versions of state, which are tracked using lineage graphs. This kind of a computation model allows Spark Streaming to recover from failures in a parallel fashion without the need for full replication (except replicating new input and periodic checkpointing to reduce recovery time), and from stragglers by running speculative copies of slow tasks. The common RDD-based storage and processing model of the Spark engine also facilitates integrating streaming, batch, and interactive computations. Like Storm+Trident, Spark Streaming provides an {\em exactly once} consistency semantics, i.e., a much weaker form of guarantee than ACID. Furthermore, the RDD-based state management model incurs high overhead for transactional workloads that require many fine-grained update operations. First, since RDDs are immutable, each update requires creating a new RDD. Second, the lineage graph also gets bigger as each operation needs to be logged. In addition, the D-Stream model also has several limitations. For example, the time interval-based batching model hinders defining fine-grained isolation units (e.g., a single tuple) or tuple-based windowing operations.


\section{Summary \& Future Directions} \label{sec:summary}

This paper has defined a simple model of transactions for stream processing. We have presented the design and implementation of a novel system called S-Store that seemlessly combines OLTP transaction processing with our streaming transaction model. We have also shown how this symbiosis can be implemented in the context of a main-memory, OLTP database system called H-Store in a straight-forward way. S-Store is shown to outperform H-Store as well as Spark Streaming and Storm on a number of streaming workloads including a leaderboard maintenance application.

Future work includes extending S-Store to operate on multiple nodes. We plan to address a number of research issues including data and workload partitioning, distributed recovery, and distributed transaction scheduling. We also plan to investigate handling of dynamic and hybrid (OLTP+stream\-ing) workloads.

\balance

\small

\bibliographystyle{abbrv}
\bibliography{sstore}

\end{document}